\RequirePackage{fix-cm}
\documentclass[smallextended]{svjour3}       
\smartqed  
\usepackage{graphicx}
\usepackage{amsmath, mathtools}
\usepackage{tabularx}
\usepackage{arydshln}
\usepackage{xcolor}
\usepackage{footnote}
\usepackage{cite}
\usepackage{multirow}
\usepackage{appendix}

\newcommand{\green}[1]{\textcolor{green}{#1}}

\usepackage{ulem} 

\definecolor{green}{rgb}{0.1, 0.5, 0.2}

\begin{document}

\title{Capacitance and Structure of Electric Double Layers:
Comparing Brownian Dynamics and Classical Density Functional Theory}



\titlerunning{Capacitance and structure of electric double layers}        

\author{ Peter   Cats 
        \and
        Ranisha S. Sitlapersad
        \and
        Wouter K. den Otter \and
        Anthony R. Thornton \and
        Ren\'e van Roij 
}


\institute{R.S. Sitlapersad, W.K. den Otter and A.R. Thornton \at
              MESA+\\
              Faculty of Engineering Technology \\
              University of Twente \\
              P.O. Box 217 \\
              7500 AE Enschede \\
              The Netherlands \\
              \email{a.r.thornton@utwente.nl}           
           \and
           P. Cats and R. van Roij \at
           Institute for Theoretical Physics \\
           Utrecht University\\
              Princetonplein 5\\
              3584 CC Utrecht\\ The Netherlands\\
              \email{p.cats@uu.nl} 
\and
        Peter Cats and Ranisha Sitlapersad contributed equally.
}

\date{Received: date / Accepted: date}

\maketitle

\begin{abstract}
We present a study of the structure and differential capacitance of electric double layers of aqueous electrolytes. We consider Electric Double Layer Capacitors (EDLC) composed of spherical cations and anions in a dielectric continuum confined between a planar cathode and anode. The model system includes steric as well as Coulombic ion-ion and ion-electrode interactions. We compare results of computationally expensive, but \lq\lq exact\rq\rq , Brownian Dynamics (BD) simulations with approximate, but cheap, calculations based on classical Density Functional Theory (DFT).    Excellent overall agreement is found for a large set of system parameters --~including variations in concentrations, ionic size- and valency-asymmetries, applied voltages, and electrode separation~-- provided the differences between the canonical ensemble of the BD simulations and the grand-canonical ensemble of DFT are properly taken into account. In particular a careful distinction is made between the differential capacitance $C_N$ at fixed number of ions and $C_\mu$ at fixed ionic chemical potential. Furthermore, we derive and exploit their thermodynamic relations. In the future these relations are also useful for comparing and contrasting experimental data with theories for supercapactitors and other systems. The quantitative agreement between simulation and theory indicates that the presented DFT is capable of accounting accurately for coupled Coulombic and packing effects. Hence it is a promising candidate to cheaply study room temperature ionic liquids at much lower dielectric constants than that of water.   

\end{abstract}

\section{Introduction}
\label{intro}

Electric double layer capacitors (EDLCs) are promising energy storage devices, in which electric energy is stored in the net ionic charge that is present in the vicinity of an electrode-electrolyte interface. In EDLCs the cathode attracts cations and repels anions and vice versa for the anode; more so the higher the applied voltage between the cathode and the anode \cite{Bindra_and_Revankar_2019}.
This energy storage mechanism leads to much higher \textit{power} densities than those of batteries; the discharge of the so-called Electric Double Layer (EDL)  of an EDLC can be much faster than the redox reactions in batteries \cite{Wang_2016, Winter_and_Brodd_2004}. However, the \textit{energy} densities of EDLCs are much lower than those of batteries \cite{Wang_2016}. 

One of the factors that contributes to the low energy density in EDLCs is the limited potential window in which conventional electrolytes are stable with respect to detrimental chemical reactions.  Conventional electrolytes in EDLCs consist of a salt (e.g. tetraethylammonium tetrafluoroborate) dissolved in a solvent (e.g. propylene carbonate or acetonitrile) \cite{Begium_2014}. In order to maximise the energy density of EDLCs, one could use alternative electrolytes with a larger potential window \cite{Liu2019}. Room temperature Ionic liquids (ILs) are potential alternatives to conventional electrolytes in EDLCs, since they can have a potential window of up to 6 V \cite{Galinski2006}; conventional electrolytes in EDLCs have a potential window of only 2.5 to 2.8 V \cite{Liu2019}. Other advantages of ILs over conventional electrolytes are their stability at high temperatures and a low vapour pressure (low volatility and non-flammability), which makes IL-based EDLCs much safer \cite{Liu2019, Salanne2018}. Modelling and theoretically understanding concentrated ILs is difficult, because of the steric repulsions at short ionic separations and the Coulombic interactions at longer ranges are simultaneously at play. While in experiments dispersion forces, polarisation, and orientation degrees of freedom often also play a role, we restrict attention to the combined effects of packing and electrostatics. In order to prepare for the challenges posed by ILs, we here focus on the parameter regime of aqueous systems.  

Several methods have been applied to investigate the electric double layer of EDLCs. On the one hand there are continuum methods, such as the mean field Gouy-Chapman-Stern (GCS) theory for point ions and classical Density Functional Theory (DFT) that can include steric effects. On the other hand there are methods in which each ionic constituent is treated explicitly, such as in Molecular Dynamics (MD) and Brownian Dynamics (BD) simulations. All have their advantages and disadvantages. GCS theory can be solved analytically, but is rather inaccurate for larger ionic concentrations and surface charges, DFT is computationally fast but is an approximate theory for a given model. MD simulations might be considered as `exact' and provide dynamics at the molecular scale but are computationally expensive.
DFT calculations \cite{Haertel_2017,Forsman_2011,Fedorov_2014,Henderson_2011,Jiang_2011,Yang_2020,Shen_2020} and MD simulations are extensively used to study  double layers in ILs and aqueous electrolytes \cite{Lanning2004, Federov2008, Feng2009, Kislenko2009, Paek2012, Si2012, Vatamanu2010, Merlet2011, Merlet2013, Yang_and_Wang, Vatamanu2011, Reed2007, Pounds2009, Vatamanu2011b,Bo_2015,Jiang_2016, Crozier_2000, Crozier_2001, Spohr_2002}.
As an alternative to the MD simulation method, BD is less accurate but computationally cheaper. In BD the explicit solvent of MD is eliminated
  by including solvent effects
      --~like friction, Brownian noise
        and the dielectric constant of the medium~--
    in an approximate way in the equations of motion of the ions \cite{AllenTildesley}.

The main aim of the paper is to explore
    whether DFT can be used to efficiently model electrolytes;
  therefore we are interested in the ability of DFT to
    quantitatively match BD simulations of a primitive model electrolyte.
To this end, we study the EDLs of an aqueous electrolyte confined between a planar anode and cathode using BD simulations and classical DFT. A given potential difference is applied between the two electrodes, which are modeled as graphene-like
electrodes in the BD simulations and as impenetrable planar walls in the DFT calculations. We compare both the ion concentration profiles
between the electrodes  (see also Refs.\cite{Gillespie_2018,Gillespie_2018_MC}) and the differential capacitance $C$.
The differential capacitance characterises the (additional) charge per (additional) applied voltage, a quantity that is experimentally measurable and often used as a characteristic of the energy storage qualities of a capacitor.  

Also it is shown that a careful distinction is required between the capacitance $C_N$ at constant number of ions and $C_\mu$ at constant ion chemical potential, where the former follows naturally from the canonical BD simulations and the latter from the grand-canonical DFT calculations.  However, they are  related and we show the connecting expressions.  We start by a detailed study of our {\it reference system}: a 1\,mol$\cdot$L$^{-1}$ 1:1 electrolyte of equal-sized ions of diameter $d=0.5$nm in the 4nm gap between two electrodes at a potential difference of 0.2V.  Then we vary the salt concentration, the ion valencies, the diameter ratio, the electrode-electrode distance, and the applied voltage.  Throughout, the whole explored 5-dimensional parameter space /design of experiments we find excellent agreement between our BD and DFT results.


\section{Model}\label{sec:model}

\begin{figure}
    \centering
    \includegraphics[width=0.8\textwidth ]{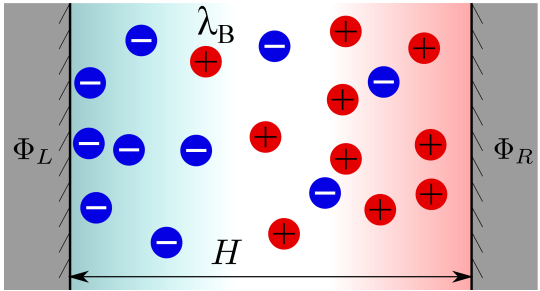}
    \caption{Illustration of the electrolyte with cations (+) and anions (-) dissolved in a dielectric medium characterized by the Bjerrum length $\lambda_B=\beta e^2/4\pi\epsilon_0\epsilon$ confined between two planar electrodes separated by a distance $H$ at which a potential $\Phi_L$ and $\Phi_R$ is applied on the left and right electrode, respectively. The gradient background color indicates the charge density.}
    \label{fig:system}
\end{figure}

We consider an aqueous electrolyte confined by two planar electrodes at fixed surface potential $\Phi_L$ and $\Phi_R$, separated by a distance $H$. The electrolyte contains  spherical cations (+) and anions (-) with a diameter $d_\pm$ and valency $Z_\pm$ dissolved in a structureless medium with dielectric constant $\varepsilon=78$ at room temperature {$T=298$ K} (see Fig.~\ref{fig:system}). The medium is fully characterized by its Bjerrum length $\lambda_B=\beta e^2/4\pi\varepsilon\varepsilon_0=0.72$ nm, where $\beta=1/k_BT$ and $\varepsilon_0$ the dielectric permittivity of free space. 
The pair potential $u_{ij}(r)$ between a pair of ions of species $i$ and $j$ separated by a distance $r$ is composed of a steric repulsions, characterized by the diameter $d$, and the Coulombic interaction, i.e.
\begin{align}\label{Eq:pair_pot}
\beta u_{ij}(r)=\beta u^{rep}_{ij}(r)+Z_iZ_j\frac{\lambda_B}{r}.
\end{align}
In the DFT calculations we describe the steric repulsions with a hard-sphere potential
\begin{align}\label{Eq:pair_pot_HS}
\beta u_{ij}^{HS}(r)=
\begin{cases}
\displaystyle \infty &r<d_{ij};\\
\displaystyle 0 &r>d_{ij},
\end{cases}
\end{align}
whereas in the BD simulations we employ the Weeks-Chandler-Andersen (WCA) pair potential
\begin{align}\label{Eq:pair_pot_WCA}
u_{ij}^{WCA}(r)=
 \begin{dcases}
 4\epsilon \left[\left(\dfrac{d_{ij}}{r}\right)^{12} - \left(\dfrac{d_{ij}}{r}\right)^6\right] + \epsilon  & \quad r < 2^{1/6} d_{ij};\\
 0  & \quad r > 2^{1/6} d_{ij}.
 \end{dcases}
\end{align} 
Here $\epsilon$ is the interaction parameter that we set to $\epsilon=k_BT$, and $d_{ij}=(d_i+d_j)/2$. Note that the WCA potential, which is just the repulsive part of the Lennard-Jones potential,  is only slighty softer for $r<d$ than the hard-sphere potential used in the DFT calculations. However, we will show remarkable agreement between the DFT and BD results in the parameter regime of study, indicating the limited sensitivity of the functional form of the repulsive interaction. 

By confining the system, we also introduce an external potential, i.e. the interaction of the ions with the fixed particles in the electrode. 
 This ion-electrode interaction is described by the WCA potential and Coulombic interactions, in both DFT and BD. In the DFT description we integrate out the in-plane dimensions, finding for the WCA part of the external potential
\begin{align}
    \beta V_{ext}^j(z<d_{wj})=2\pi\rho_w\epsilon_w d_{wj}^2 &\left[\frac{2^{4/3}}{5}+\frac{2}{5}\left(\frac{d_{wj}}{z}\right)^{10}-\left(\frac{d_{wj}}{z}\right)^4\right. \nonumber \\
    &\left.-\frac{1}{2}\left(2^{1/3}-\left(\frac{z}{d_{wj}}\right)^2\right)\right]
\end{align}
where $\rho_w$ is the surface density of wall particles in number of particles per unit area; $\epsilon_w$ the interaction strength between the wall particles and ions, for which we take $\epsilon_w=\epsilon=k_BT$; the contact distance is $d_{wj}=(d_w+d_j)/2$ with $d_w$ and $d_j$ the diameter of the wall particles and ions of species $j$, respectively. For $z>d_{wj}$ the external potential vanishes. Note that the second electrode is described by the same interaction potential with $z$ replaced by $H-z$. Our primitive model captures the key features of an electrolyte,
    within the practical conditions posed by DFT and BD respectively,
  thereby enabling a quantitative comparison between both methods.
For simplicity, we use the same diameter for all ions and electrode particles in the reference system, namely $d= 0.5$\;nm.


\section{Method}\label{sec:method}

\subsection{Brownian Dynamics}\label{sec:method_BD}

\begin{figure}
\centering
  \includegraphics[width = 0.75\textwidth]{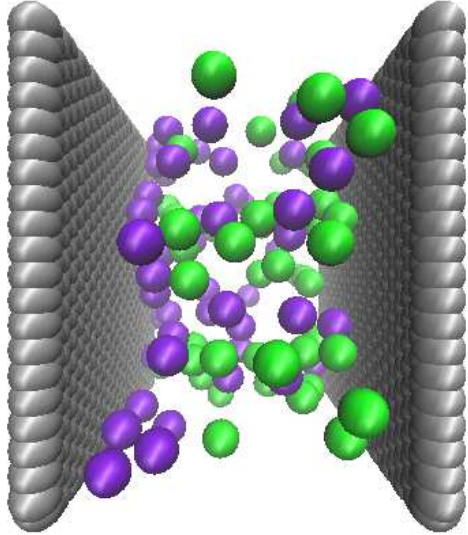}
\caption{
  Simulation snapshot of a system containing 51 ion pairs,
      corresponding to a reservoir salt concentration of
        1\,mol$\cdot$L$^{-1}$,
    between two graphene-like electrodes with a separation of $H = 4$ nm.
The potential difference between the electrodes is 0.2\,V,
  with the electrode on the right at a higher potential
    than the electrode on the left.
The valencies of the cations (purple) and anions (green)
    are $Z_+ = 1$ and $Z_- = -1$, respectively,
  and both have the same diameter $d = 0.5$ nm.
\label{fig:snapshot} 
}
\end{figure}

We focus here on the description of the reference system used in the BD simulations,
    as illustrated in Fig.\,\ref{fig:snapshot};
  detailed information on the variations to this system are provided in Section \ref{sec:results}.
The simulations were performed in LAMMPS \cite{LAMMPS}. 
The salt concentration in the slit between the two electrodes is set to what it would have been if the slit was in equilibrium with a reservoir, with a salt concentration of 1\,mol$\cdot$L$^{-1}$ at 0V. The salt concentration in the slit was determined by DFT calculations, which corresponds to an equal amount of anions and cations, $N_+ = N_- = 51$. 
As described in Section \ref{sec:model}, the pairwise excluded volume interactions are described by a WCA potential. The ions have valencies of $Z_+ = +1$ and $Z_- = -1$, while the variable charges of the wall particles are determined by the Constant Potential Method (see below). 
The long-ranged Coulombic interactions are evaluated using the particle-particle-particle-mesh method (PPPM) \cite{PPPM}, with a cut-off distance of 12 nm and a relative accuracy of $10^{-6}$ in the forces. A correction term allows application of this 3D Ewald summation technique to the current slab geometry \cite{Yeh_and_Berkowitz,Ballenegger}. In the BD simulation the solvent is implicit and accounted for in the equation of motion. That is, the second order Langevin equation of motion of the ions reads as
\begin{equation}\label{Eq:langevin}
    m \ddot{\mathbf{x}}_i = - \xi \dot{\mathbf{x}}_i - \frac{ \partial U }{ \partial \mathbf{x}_i } + \mathbf{f}_i(t),
\end{equation}
where $\mathbf{x}_i$ is the position of the $i^\mathrm{th}$ particle, $m = 50$ atomic mass unit (a.m.u.) denotes the mass of the ions (equal for all ions), $\xi$ the friction constant, $U$ the total potential energy, and $\mathbf{f}_i$ the fluctuating Brownian force on the particle. Using the Stokes-Einstein equation $\xi = 3 \pi \eta d$ yields $\xi = 2.2$\,a.m.u.$\cdot$fs$^{-1}$ in water with $\eta_{water} = 8 \times 10^{-4}$\,Pa$\cdot$s. The stochastic force, with vanishing mean and devoid of correlations in time (Markovian) and across particles, obeys the fluctuation-dissipation theorem,
\begin{equation}
    \left\langle \mathbf{f}_i(t) \otimes \mathbf{f}_j(t^\prime) \right\rangle = 2 \xi k_B T \delta_{ij} \delta( t - t^\prime) \mathbf{1},
\end{equation}
where $\delta_{ij}$ is the Kronecker delta, $\delta( t - t^\prime )$ the Dirac delta and the angular brackets denote an average. This equation of motion is integrated in LAMMPS by combining the velocity-Verlet scheme
  \cite{AllenTildesley,FrenkelSmit},
using a time step of 5\,fs, with the Langevin option.
Simulations were initiated by placing the ions, stacked into a simple cubic crystal lattice, in the slit.
A simulation typically lasted for 200\,ns,
  requiring about 2-3 days on 32 cores in parallel,
with the first 10\% serving as equilibration phase and the remainder as production run.

The two parallel electrodes each consist of one graphene-like layer of 960 particles,
covering an area of $A = 100.6\,d^2$, see Fig.\,\ref{fig:snapshot}. 
The inter-particle bond lengths in these hexagonal layers are taken as the usual carbon-carbon distance, 0.142\,nm\,$ = 0.284d$. The distance between the electrodes, as measured between the centres of the constituent particles, is $H = 8d$.
All wall particles are frozen,
    i.e. are excluded from the equation of motion,
  because the elimination of their rapid vibrations
      permits the use of a larger time step.
Their charges are calculated using the Constant Potential Method (CPM) by imposing a constant voltage difference of $\Psi = 0.2$\,V between the two walls \cite{Siepmann_1995,Reed2007}.
We use the implementation provided by Yang et al. \cite{Yang_and_Wang} which can be used as a plug-in for LAMMPS.
In brief, the CPM determines the charges of all wall particles at every simulation step by solving the linear set of equations that determines the potential at every wall particle, given the positions of all wall particles and ions. As the implementation of CPM for LAMMPS \cite{Yang_and_Wang} does not take the relative permittivity of the medium into account, the charges of all ions were multiplied by $1/\sqrt{\epsilon}$ and the potentials on the walls were multiplied with $\sqrt{\epsilon}$ to reach the same effect - all charges reported below are corrected
to refer to an aqueous medium with $\epsilon = 78$.
A side effect
    to this pragmatic inclusion of the solvent's dielectric constant
  is that not only ion-ion and ion-electrode Coulomb forces are scaled,
    but so are the Coulomb forces between electrode atoms;
  the latter is of no consequence, however,
    since these atoms are immobilized in the simulation.
A brief comparison with the fixed charge method, 
    in which all charges are permanently fixed,
  is provided in Appendix \ref{App:CMP_vs_FCP}
    for the system  at zero voltage.
The system is periodically repeated in space, with the box lengths in the two directions parallel to the walls dictated by the geometry of the lattice and the height perpendicular to the walls taken as three times the width of the slit.
Adding additional layers to the electrodes,
    creating thin slabs of graphite,
  does not significantly affect the ion density profiles
    nor the average total charge of the electrodes;
  the differences are within the accuracy of the calculation.


\subsection{Density Functional Theory}\label{sec:DFT}

The starting point of classical DFT is the grand potential functional $\Omega$ of the density profiles $\rho_j(\mathbf{r})$ \cite{Evans_1979}, which reads in our case
\begin{align}
\begin{split}
\Omega [ \{\rho\} ]
&=\mathcal{F}_{id}[\{\rho\}]+\mathcal{F}_{ex}^{HS}[\{\rho\}]+\mathcal{F}_{ex}^{ES}[\{\rho\}]
\\
&-\sum_{j=\pm} \int \mathrm{d}\mathbf{r} \rho_j(\mathbf{r})\left[\mu_j-V^j_{ext}(\mathbf{r})\right],
\end{split}
\end{align}
where $\mathcal{F}_{id}$ is the intrinsic Helmholtz free energy functional of the ideal gas, $\mathcal{F}_{ex}^{HS}$ the excess (over-ideal) Helmholtz free energy functional that deals with the hard-sphere interactions, $\mathcal{F}_{ex}^{ES}$ the excess Helmholtz free energy functional that deals with the electrostatic interactions, $\mu_j$ the chemical potential of species $j$, and $\rho_j(\mathbf{r})$ the local density of species $j$. Here, ${\mathcal{F}=\mathcal{F}_{id}+\mathcal{F}_{ex}^{HS}+\mathcal{F}_{ex}^{ES}}$ is an intrinsic property of the system which depends on the temperature and the interparticle interactions, but not on $\mu_j-V_{ext}^j(\mathbf{r})$. This grand potential functional has the property that it is minimized for a given $\mu_j-V^j_{ext}(\mathbf{r})$ by the equilibrium density profile $\rho_{j,0}(\mathbf{r})$, i.e. $\left. \delta\Omega/\delta\rho_j\right|_{\rho_{j,0}}=0$, at which it equals the actual grand potential $\Omega$ introduced later in section \ref{sec:Thermodynamics}. Minimizing the grand potential functional w.r.t. the density profiles results in the Euler-Lagrange equations
\begin{align}\label{Eq:EL}
\left.\frac{\delta \mathcal{F}[\{\rho\}]}{\delta \rho_j(\mathbf{r})}\right|_{\rho_{j,0}}=\mu_j-V^j_{ext}(\mathbf{r}).
\end{align}
Therefore, once an explicit form of $\mathcal{F}$ is constructed, one can find the equilibrium density profiles $\{\rho_0\}$ by solving Eq.~\eqref{Eq:EL}.
Although the ideal Helmholtz free energy functional is known exactly and given by
\begin{align}
\beta\mathcal{F}_{id}[\{\rho\}]=\sum_j\int \mathrm{d}\mathbf{r}\rho_j(\mathbf{r})\left[ \ln
\Lambda_j^3\rho_j(\mathbf{r})-1\right],
\end{align}
with $\Lambda_j$ the thermal wavelength, the excess functional hinges on approximations. One excellent approximation for the hard-sphere functional $\mathcal{F}_{ex}^{HS}$ has been developed and goes by the name Fundamental Measure Theory (FMT) \cite{Tarazona_85,Rosenfeld_89,Roth_2010}, of which we apply the White-Bear II version \cite{FMT_WBII}. The functional that deals with the electrostatics $\mathcal{F}_{ex}^{ES}$ is in general more difficult due to the long-range nature of the interactions \cite{Haertel_2017}. However, we use the functional based upon the Mean-Spherical-Approximation (MSA), which for the restrictive primitive model (RPM) reads \cite{MSA}
\begin{align}\label{Eq:F_ES}
\beta\mathcal{F}_{ex}^{ES}&[\{\rho\}]\approx\beta\mathcal{F}_{ex}^{MSA}[\{\rho\}]=-\frac{1}{2}\int\mathrm{d}\mathbf{r}\int\mathrm{d}\mathbf{r'}q(\mathbf{r})c^{MSA}(|\mathbf{r}-\mathbf{r'}|;\rho_r)q(\mathbf{r'}),
\end{align}
where $c^{MSA}$ is given by \cite{Waisman_1970}
\begin{align}\label{Eq:cMSA}
c^{MSA}(r)=
\begin{cases}
\displaystyle \frac{\lambda_B}{r}\frac{r(r-2D)}{D^2} & r \leq d;\\
\displaystyle -\frac{\lambda_B}{r} & r >d,\\
\end{cases}
\end{align}
and the charge density is defined by $q(\mathbf{r})=\sum_j Z_j\rho_j(\mathbf{r})$. The electrostatic potential makes its entrance upon writing Eq.~\eqref{Eq:F_ES} as the sum of the mean-field electrostatic free energy and MSA corrections, details of which can be found in Refs.~\cite{MSA,Haertel_2017,Haertel_2015}. The electrostatic potential is consequently determined by the Poisson equation, where the constant surface potential is enforced as a boundary condition.

The parameter $D=d+1/\Gamma$ is a length scale that results from MSA, where ${\Gamma=(\sqrt{1+2d\kappa}-1)/2d}$ with ${\kappa=\sqrt{8\pi\lambda_B\rho_r}}$ the inverse Debye length and $2\rho_r$ the total ion concentration in the reservoir at a given chemical potential $\mu$. For the expression of $c^{MSA}$ beyond the RPM, we refer to \cite{Blum_Ros,Hiroike}.  Note that $c^{MSA}$ depends on the reservoir concentration $\rho_r$ through the parameter $D$ and therefore it also depends on the chemical potential $\mu$. Due to the approximation for the Helmholtz free energy functional $\mathcal{F}^{ES}_{ex}$, it now also depends on $\mu$, which it formally should not depend on. As a result, the Maxwell relation introduced in the next section does not hold exactly and we have two \lq routes\rq\ to calculate the adsorption $\Gamma$ (see Appendix~\ref{App:Gam_incon}). When we need to calculate the adsorption $\Gamma$ we use the route Eq.\eqref{Eq:Gamma} given  in the next section for the remainder of the manuscript. 

As a final note we mention that our DFT calculations assume planar symmetry in which the in-plane coordinates can again be integrated out, leaving the normal coordinate $z$ perpendicular to the electrodes as the only spatial variable in the numerical calculations.


\section{Thermodynamics\label{sec:Thermodynamics}}

We will treat the above-mentioned model theoretically via DFT, and by BD simulations. However, firstly we discuss the thermodynamics of both methods.

\begin{figure}
    \centering
    \includegraphics{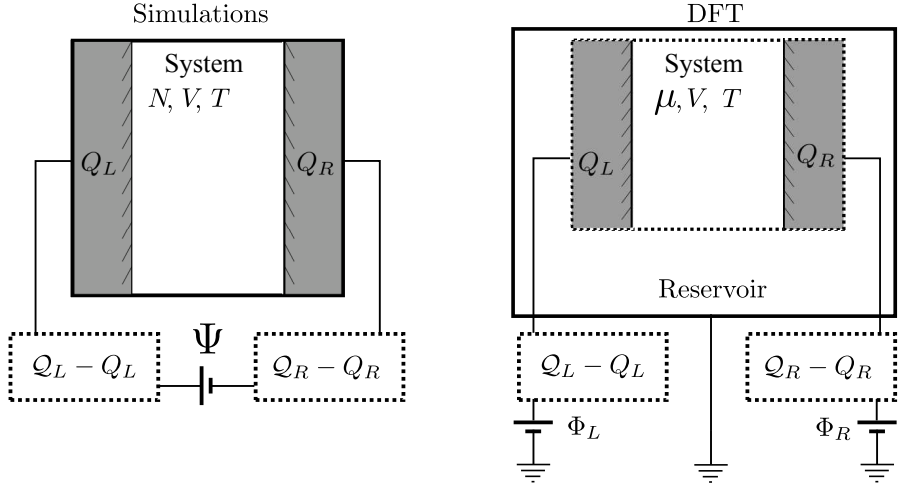}
    \caption{Illustration of the thermodynamic ensembles applicable to the BD simulations (left) and DFT (right). The simulations are performed with a fixed number of particles $N$ in a fixed volume $V$ at a fixed temperature $T$, while the DFT calculations employ a fixed chemical potential $\mu$ or reservoir concentration $\rho_r$. In the simulations the potential difference between the electrodes is fixed at $\Psi$, while in the DFT calculations the potentials of both electrodes relative to the reservoir are fixed at  $\Phi_{L/R}$ for the left and right electrode, respectively. In both cases, the electrodes can exchange charge with charge reservoirs to maintain the imposed potential difference(s).}
    \label{fig:system_thermo}
\end{figure}

Let us start by considering the ensemble of the BD simulations, which is a closed system with a fixed number of $N_+$ cations and $N_-$ anions in a volume $V$ at temperature $T$ confined between two planar electrodes of equal area $A$ separated by a distance $H$ and held at a surface potential difference $\Psi$ (see Fig.~\ref{fig:system_thermo}). The system, i.e. electrolyte and the electrodes, together with the charge reservoirs is charge neutral $\sum_{j=\pm}eZ_jN_j+\mathcal{Q}_L+\mathcal{Q}_R=0$. Applying a potential difference $\Psi$ creates an electric field across the system. Since the ions in the system are mobile they will respond to this electric field, and because the electrolyte is confined they will create a charge density near both electrodes. These electrodes are connected to charge reservoirs with which they can exchange charges, such that the charge on the electrode is balanced with the charge density in the electrolyte.  In other words, the charge in the electrolyte together with the average charge on the electrodes must vanish, i.e. $\sum_{j=\pm}eZ_jN_j+ Q_L+ Q_R=0$, where $Q_{L/R}$ denotes the average charge.
The corresponding thermodynamic potential for this system is the free energy $F(N_+,N_-,V,T,\Psi,A,H)$, for which the differential form reads
\begin{align}\label{Eq:dF}
    \mathrm{d}F=-S\mathrm{d}T-p\mathrm{d}V+\sum_{j=\pm}\mu_j \mathrm{d}N_j-(Q_R-Q_L)\mathrm{d}\Psi+\gamma\mathrm{d}A-f\mathrm{d}H,
\end{align}
with $S$ the entropy, $p$ the pressure, $\mu_\pm$ the chemical potential of the cations and anions, respectively, $A$ the surface area, and $\gamma$ is the total surface tension (which has contributions from the electrode-electrolyte and in the case of EDL-overlap also from electrode-electrode interactions).   
We also introduced the force $f$ between the two planar electrodes, were $f/A$ is also referred to as the disjoining pressure. We will call $F$ the Helmholtz free energy, even though it is only a Helmholtz free energy for the ionic species while it is actually a grand-canonical potential for the charge carriers in the electrodes.   

The DFT calculations, on the other hand, are performed at constant chemical potential $\mu_\pm$ instead of constant number of ions $N_\pm$. This implicitly means that the system can {freely} exchange ions with an ion reservoir. Also, both the electrode potentials are defined w.r.t. a grounded reservoir (see Fig.~\ref{fig:system_thermo}), i.e. the two electrodes are connected to separate charge reservoirs that are independently held at a constant potential. Hence, the potential $\Phi_L$ and $\Phi_R$ on the left and right electrode, respectively, generate an independent electric field, not only between the electrodes but also between the electrodes and the reservoir. The ions both within the system and in the reservoir respond to this electric field. The role of the charge reservoirs is the same in both ensembles. Since global charge neutrality of the system plus reservoirs still holds, one finds in equilibrium the system charge neutrality condition  $\sum_{j=\pm}eZ_j\langle N_j\rangle+\langle Q_L\rangle+\langle Q_R\rangle=0$.
 The corresponding thermodynamic potential is the grand potential $\Omega(\mu_\pm,V,T,\Phi_L,\Phi_R,A,H)$ with differential 
\begin{align}\label{Eq:dOm}
    \mathrm{d}\Omega=-S\mathrm{d}T-p\mathrm{d}V-\sum_j N_j \mathrm{d}\mu_j-Q_L\mathrm{d}\Phi_L-Q_R\mathrm{d}\Phi_R+\gamma \mathrm{d}A-f\mathrm{d}H.
\end{align}
The distinction between $F(N_\pm,V,T,\Psi,A,H)$ and $\Omega(\mu_\pm,V,T,\Phi_L,\Phi_R,A,H)$ is crucial when comparing results form DFT (constant chemical potential) with BD simulations (constant number of ions). 

For macroscopically large systems we can use volumetric and areal extensivity arguments to write
$\Omega=-p(\mu_+,\mu_-)V+\gamma(\mu_+,\mu_-,\Psi_L,\Psi_R,H)A$, where we drop the $T$ dependence for convenience as we keep the temperature fixed throughout. Combining the resulting differential $d\Omega=-pdV-Vdp +\gamma dA + A d\gamma$ with  Eq.(\ref{Eq:dOm}) gives the Gibbs-Duhem equation for the volumetric terms and, for $dH=0$, the Lipmann equation 
\begin{align}\label{Eq:Lipmann}
    \mathrm{d}\gamma=-\sigma_L\mathrm{d}\Phi_L-\sigma_R\mathrm{d}\Phi_R-\sum_j\Gamma_j\mathrm{d}\mu_j,
\end{align}
where $\sigma_{L/R}=Q_{L/R}/A$ denotes the surface charge density and $\Gamma_j$ the adsorption of ions of species $j$ onto both electrodes defined by 
\begin{align}\label{Eq:Gamma}
    \Gamma_j=\int_0^H \mathrm{d}z \left(\rho_j(z)-\rho_{j,r}\right).
\end{align}
Here, $z$ denotes the coordinate describing the distance perpendicular to the parallel electrodes, $\rho_j(z)$ the local number density of species $j$ at position $z$, and $\rho_{j,r}$ the reservoir concentration of species $j$ which is dictated by the chemical potentials $\mu_\pm$. See Appendix~\ref{App:Thermo} for a more detailed derivation.

The main observable that we focus on in this manuscript is the differential capacitance (per unit area), which can either be obtained at constant number of particles $C_N$  or at constant chemical potential $C_\mu$, i.e.
\begin{align}\label{Eq:CapmuN}
    C_N&= \left(\frac{\partial\sigma}{\partial \Psi}\right)_N=-\frac{1}{A}\left(\frac{\partial^2 F}{\partial \Psi^2}\right)_N, &     C_\mu&=\left(\frac{\partial \sigma}{\partial \Psi}\right)_\mu=-\frac{1}{A}\left(\frac{\partial^2 \Omega}{\partial \Psi^2}\right)_\mu,
\end{align}
where, for simplicity, we consider the RPM where $N_+=N_-\equiv N$ and $\mathrm{d}\mu_+=\mathrm{d\mu_-}\equiv \mathrm{d}\mu/2$, which allows us to write  $\Gamma=(\Gamma_++\Gamma_-)/2$. On top of that, we also apply the same (but opposite) potential on both electrodes within the $\Omega$ ensemble, i.e. $\Phi_R=-\Phi_L\equiv \Psi/2$, which leads to the same (but opposite) surface charge  $\sigma_L=-\sigma_R\equiv \sigma$ on both electrodes. 

Interestingly, $C_\mu$ and $C_N$ are related via expressions that are very similar to those between the constant-volume and constant-pressure heat capacities, namely (see Appendix \ref{App:Thermo})
\begin{align}\label{Eq:CmuCN}
   C_\mu-C_N&=\frac{\alpha^2}{\chi_\Psi}\geq 0
   && \mathrm{and} &&
   \frac{C_\mu}{C_N}=\frac{\chi_\Psi}{\chi_\sigma}\geq 1,
\end{align}
where $\alpha$ can be taken equal either to $\alpha_\mu$ or $\alpha_\Psi$ defined by  
\begin{align}\label{Eq:alphas}
    \alpha_\mu\equiv\left(\frac{\partial \Gamma}{\partial \Psi}\right)_\mu=\left(\frac{\partial \sigma}{\partial \mu}\right)_\Psi\equiv\alpha_\Psi.
\end{align}
In an exact theory the Maxwell relation $\alpha_\mu=\alpha_\Psi$ is satisfied identically, however our excess free-energy functional is approximate and does not identically satisfy the Maxwell relation. As a consequence our conversion between $C_N$ and $C_\mu$ depends on the choice for $\alpha$, and thus leads to an inconsistency. However, the numerical differences are limited, as we will see in section~\ref{sec:Cap}.   We also defined 
\begin{align}
    \chi_\sigma=\left(\frac{\partial N}{\partial \mu}\right)_\sigma
    && \mathrm{and} &&
    \chi_\Psi=\left(\frac{\partial N}{\partial \mu}\right)_\Psi,
\end{align}
     which resemble (osmotic) compressibilities of the ions at constant $\sigma$ and $\Psi$, respectively. 
     The relations in this section allow us to compare and convert the capacitances at constant $N$ (natural to BD simulation) and at constant $\mu$ (natural to DFT).

In the BD simulations,
  the differential capacitances were determined by three routes.
Running a set of equilibrium simulations
    at a range of potential differences $\Psi$ between the electrodes
  yielded the mean total charge difference between the electrodes,
  $\langle Q \rangle_{N,\Psi}
    = (1/2) \langle Q_R - Q_L \rangle_{N,\Psi}$.
The angular brackets denote
    the canonical ensemble average at the indicated potential difference,
  which is evaluated in simulations as a time-average
    \cite{AllenTildesley,FrenkelSmit}.
The differential capacitance at constant numbers of ions,
    see Eq.~\ref{Eq:CapmuN},
  is obtained
    by numerically differentiating the surface charge
      with respect to the potential difference, 
\begin{align}
    C_N^\Delta(N, \Psi )
  =
    \frac{
      \langle Q \rangle_{ N, \Psi + \Delta \Psi }
      -
      \langle Q \rangle_{ N, \Psi - \Delta \Psi }
    }{
      2 A \Delta \Psi
    }
\end{align}
  using the central difference formula.
 Alternatively, the same differential capacitance is extracted from the thermal fluctuations of the wall charge around the average,
         $\delta Q = Q - \langle Q \rangle_{N,\Psi}$,
       over a single equilibrium simulation \cite{Limmer2013,Scalfi_2019},
\begin{align}\label{Eq:CN_fluc}
    C_N^\delta( N,\Psi )
  &=
    \frac{
      \left\langle \left( \delta Q \right)^2 \right\rangle_{N,\Psi}
    }{ k_B T A } + C_0,
\end{align}
The intrinsic capacitance $C_0$, which is a constant independent of $\Psi$ and $N$, accounts for the thermal fluctuations of the atomic charges around the idealized constant-potential charges calculated by CPM. 
The numerical value of $C_0$ is obtained by the fitting procedure discussed
    in Appendix \ref{App:calc_cap}.

The differential capacitance at constant chemical potential, 
    see Eq.~\ref{Eq:CapmuN},
  is obtained from the BD simulations as
\begin{align}\label{Eq:CDmu}
    C_\mu^\Delta(N, \Psi )
  =
    \frac{
      \langle Q \rangle_{ N( \mu, \Psi + \Delta \Psi ), \Psi + \Delta \Psi }
      -
      \langle Q \rangle_ {N( \mu, \Psi - \Delta \Psi ), \Psi - \Delta \Psi }
    }{
      2A \Delta \Psi
    },
\end{align}
  where each simulation was preceded by
    a DFT calculation to establish the required numbers of ions $N(\mu,\Psi)$
      under the prevailing conditions.
The same finite difference relations were used to calculate the capacitance by DFT.
The calculations of $C_N$ via Eq.\eqref{Eq:alphas}
  will be referred to as $C^{\alpha_\mu}_N$ and $C^{\alpha_\Psi}_N$,
    when using $\alpha_\mu$ and $\alpha_\Psi$, respectively.


\section{Results}\label{sec:results}

In this section,
  the density profiles and the differential capacitances
    obtained with BD and DFT are presented and compared.
The reference system will be considered first,
  followed by an exploration of the impacts of its various parameters
    by varying them one by one.
Various authors have previously used DFT and MD
  to study systems similar to the systems discussed below \cite{Haertel_2015,MSA,Haertel_2017,Gillespie_2003,Roth_2016,Crozier_2000,Crozier_2001,Spohr_2002}.
    Although those systems are not identical to the systems we consider,
      they do show very similar density profiles.


\subsection{Reference system}

\begin{figure}
    \centering
    \includegraphics[width = 0.75\textwidth]{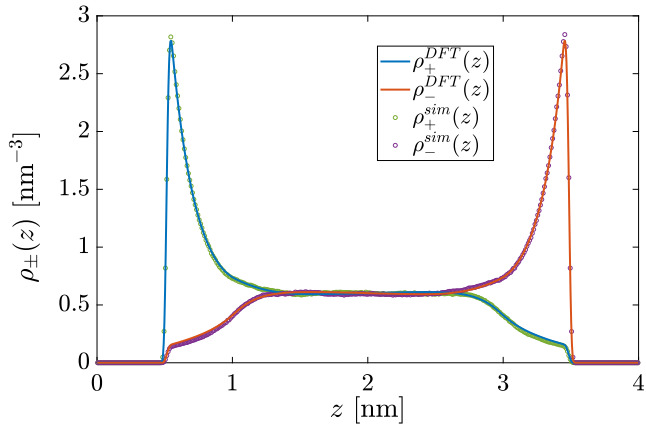}
        \caption{
        The reference system, where the electrodes are separated by $H=4$ nm and are held at a potential $\Phi_L=-0.1$ V and $\Phi_R=0.1$ V ($\Psi=0.2$ V), for the left and right electrode respectively.
        The electrolyte consists of monovalent ions with radius $d=0.5$\,nm
        at a reservoir concentration  $\rho_r=1$\,mol$\cdot$L$^{-1}$ in the reservoir,
          resulting in 51 ion pairs in the BD simulation.}
    \label{fig:reference}
\end{figure}

In the reference system, the electrodes are separated by $H = 4 \, {\rm nm}$ while the applied electrode potentials are $ \Phi_L=-0.1 \, {\rm V}$ and $\Phi_R = +0.1 \, {\rm V}$ in DFT and $\Psi=0.2$ V in the BD simulations. The reservoir salt concentration is $\rho_r=1$\,mol$\cdot$L$^{-1}$, \textit{i.e.} cations and anions have identical concentrations of 1\,mol$\cdot$L$^{-1}$, which corresponds to 51 ion pairs in the BD system with electrode surface areas of $A=25.1$\,nm$^2$. The cations and anions have the same size $d_+ = d_- =0.5 \, {\rm nm}$, and are monovalent $Z_+ = -Z_- = 1$, see Fig.~\ref{fig:snapshot}. The results from DFT and the simulations for this reference system are presented in Fig.~\ref{fig:reference}. Applying a negative (positive) surface potential on the left (right) electrodes causes a negative (positive) surface charge, attracting positively (negatively) charged ions and repelling negative (positively) charged ions. Away from the walls, the concentrations of both ions level off to a flat density profile.
The agreement between DFT and BD is excellent.

\begin{figure}
\centering
    \includegraphics[width = 0.425\textwidth]{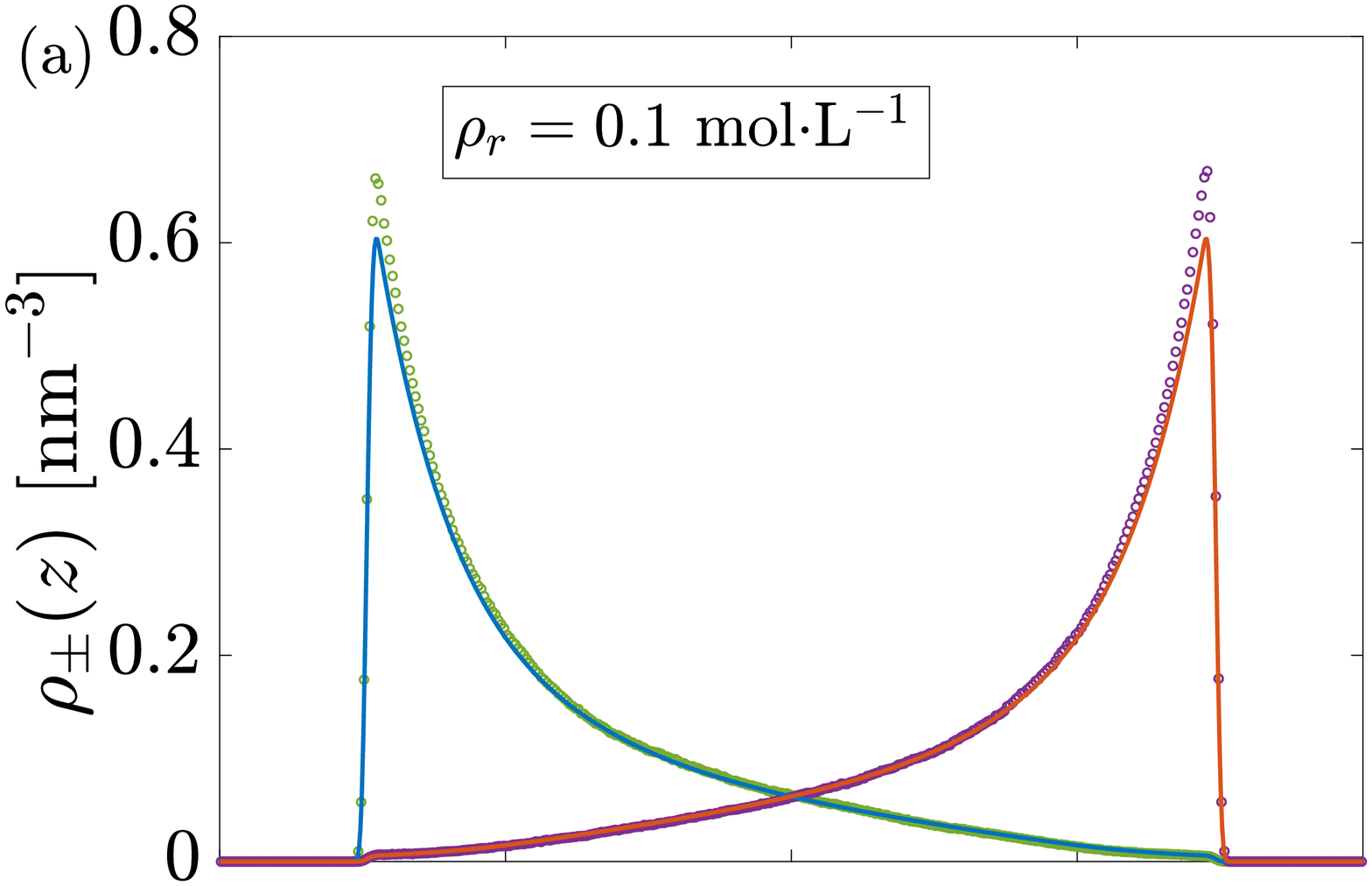}
    \includegraphics[width = 0.425\textwidth]{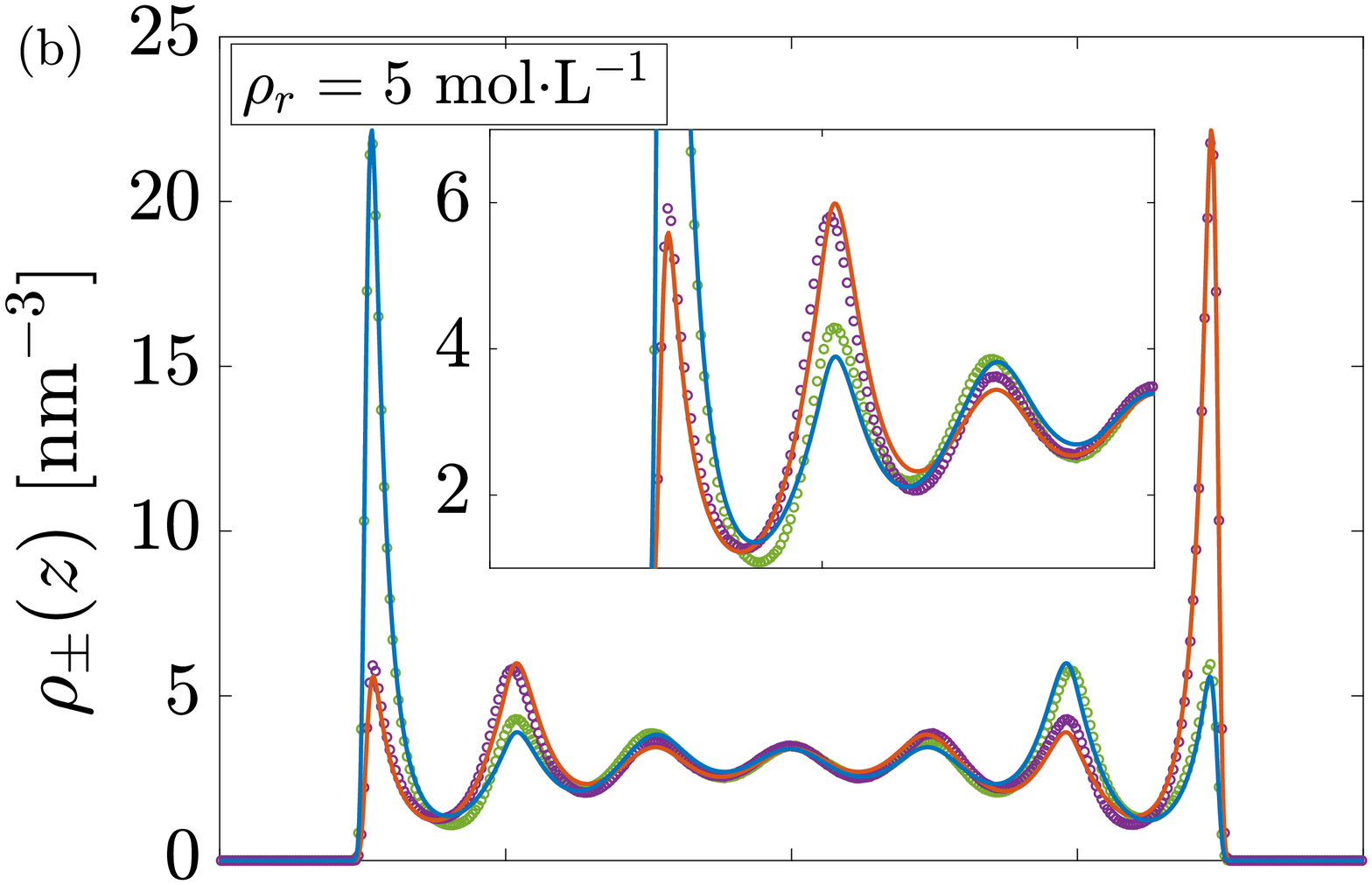}\\
    \includegraphics[width = 0.425\textwidth]{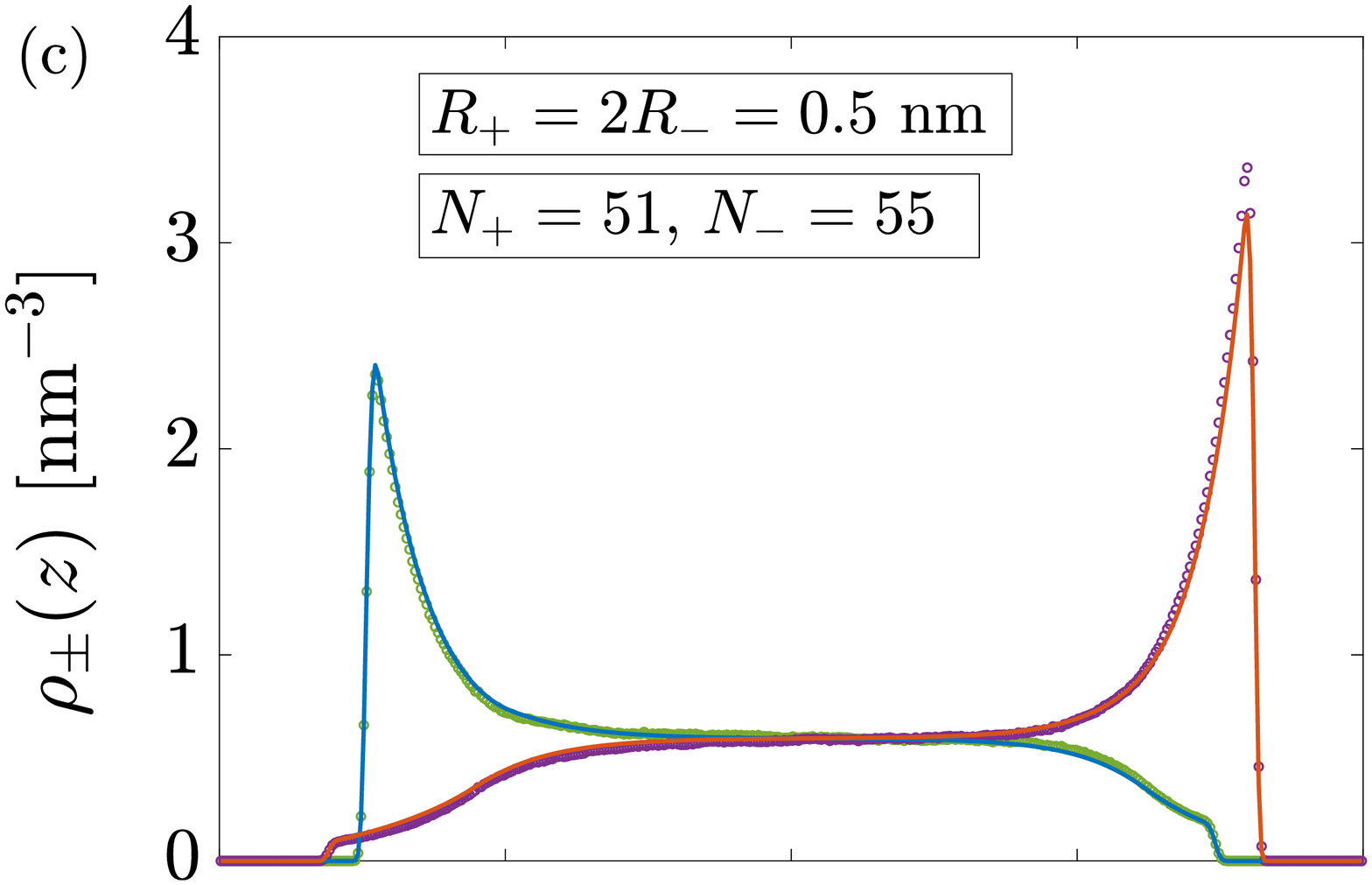}
    \includegraphics[width = 0.425\textwidth]{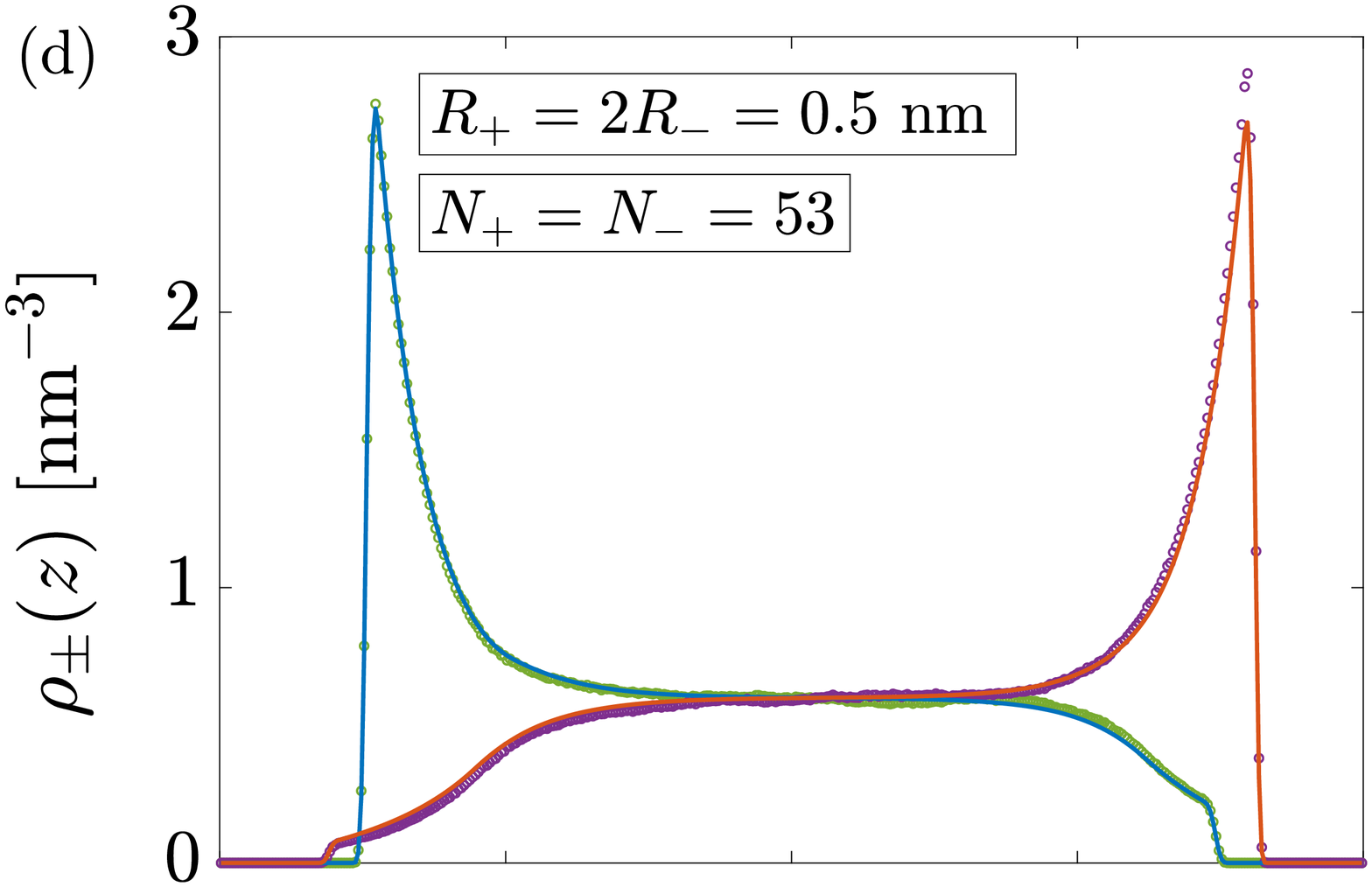}\\
    \includegraphics[width = 0.425\textwidth]{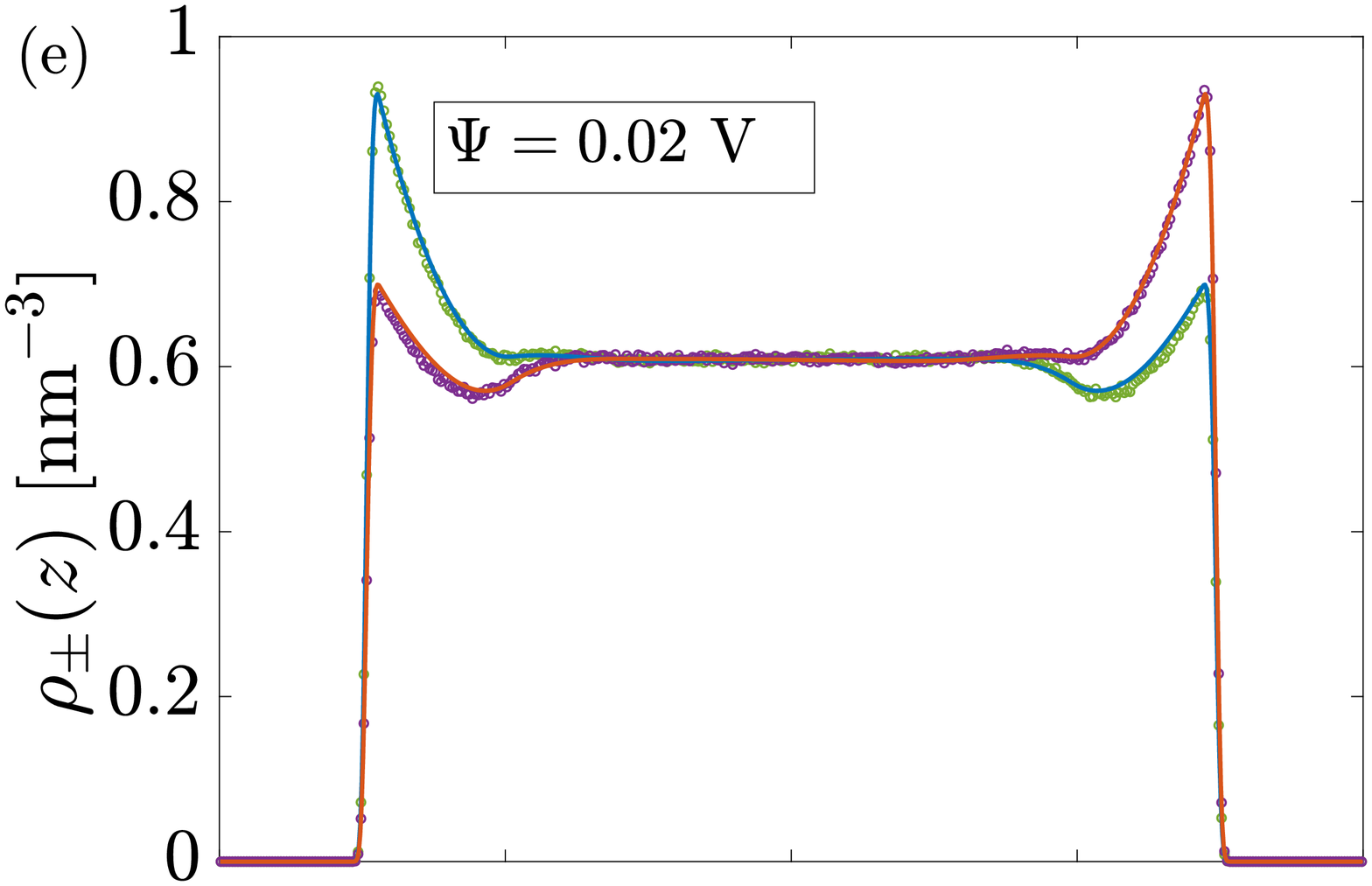}
    \includegraphics[width = 0.425\textwidth]{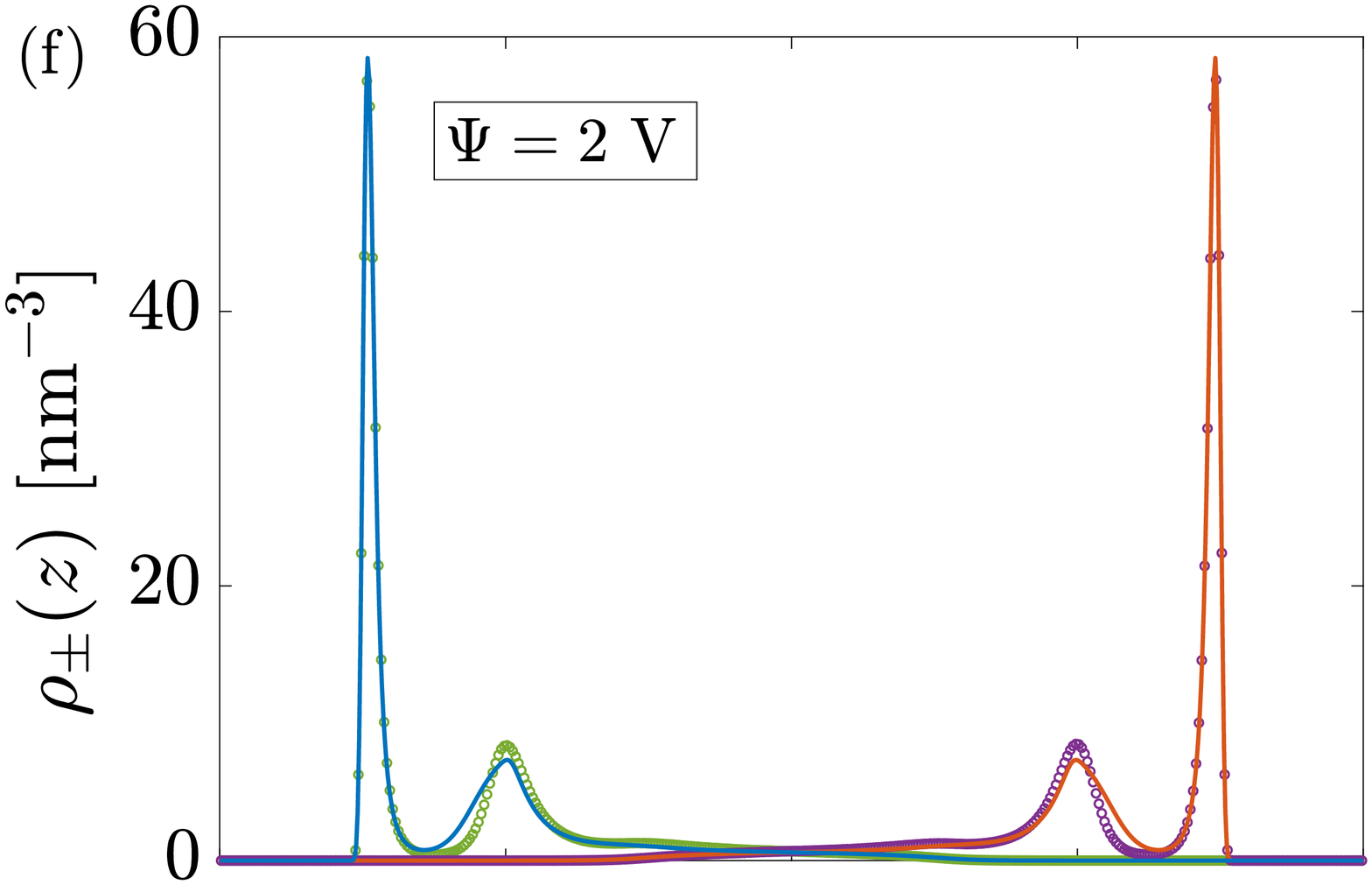}\\
    \includegraphics[width = 0.425\textwidth]{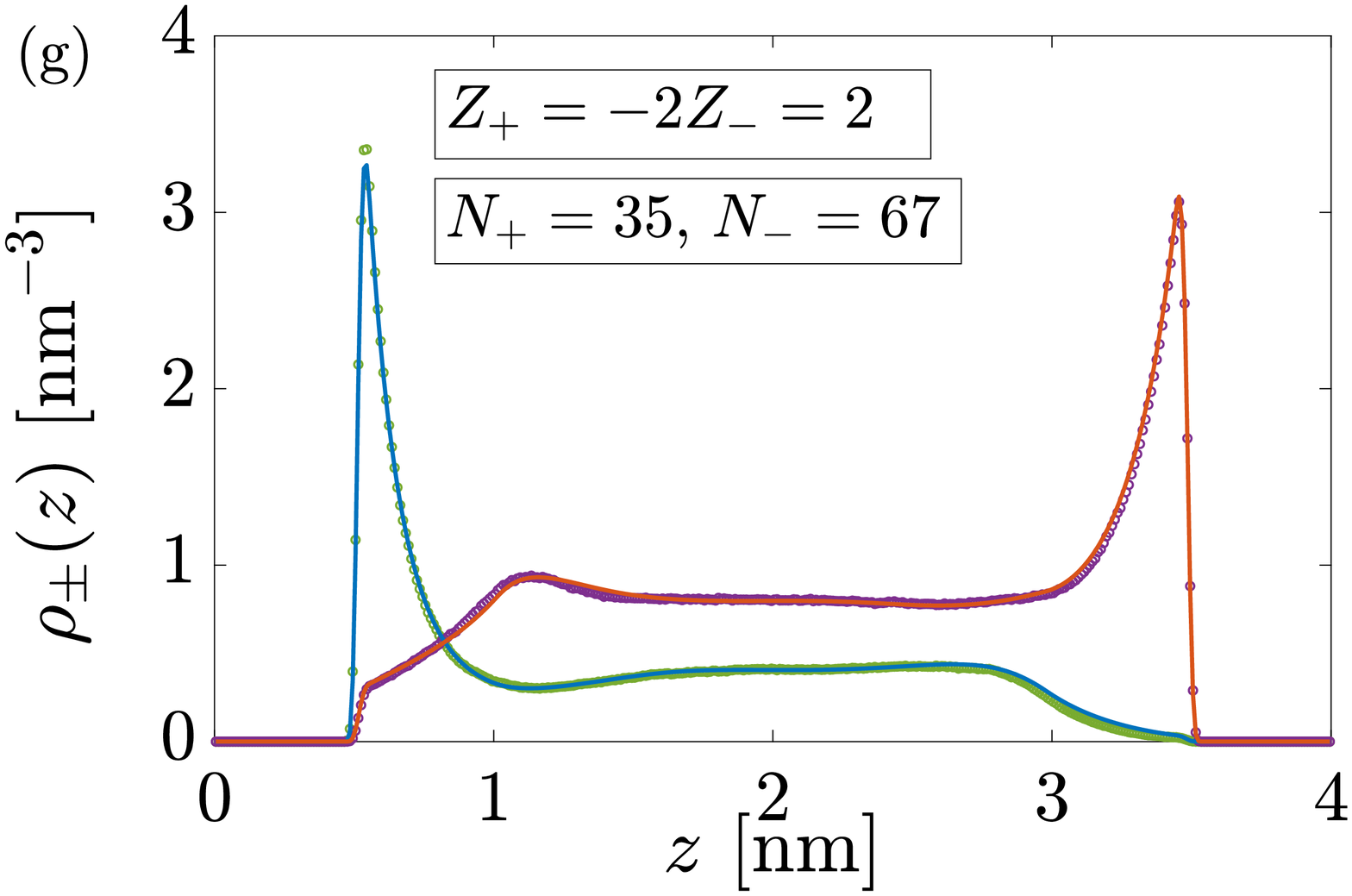}
    \includegraphics[width = 0.425\textwidth]{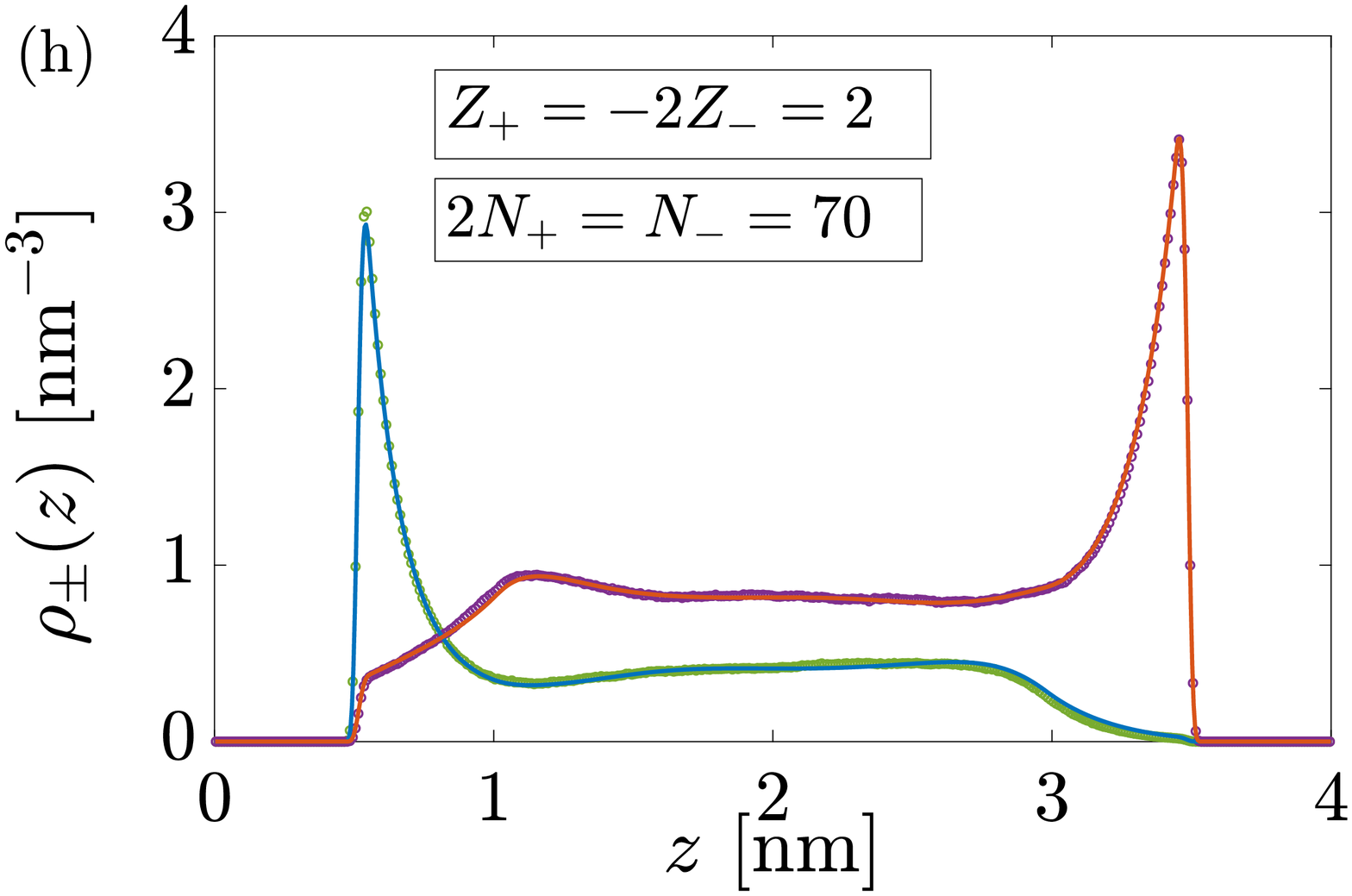}\\
    \includegraphics[width = 0.425\textwidth]{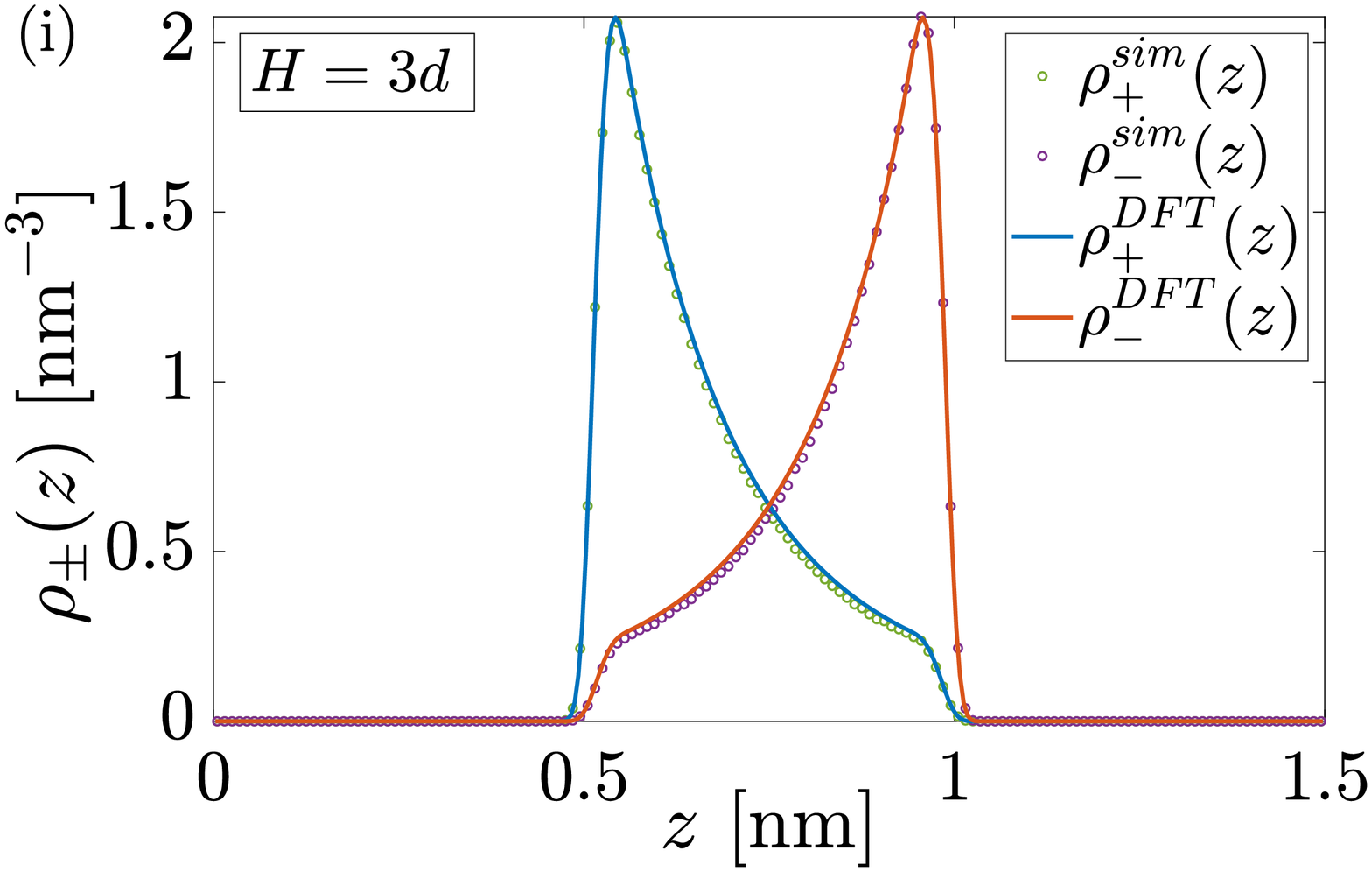}
    \includegraphics[width = 0.425\textwidth]{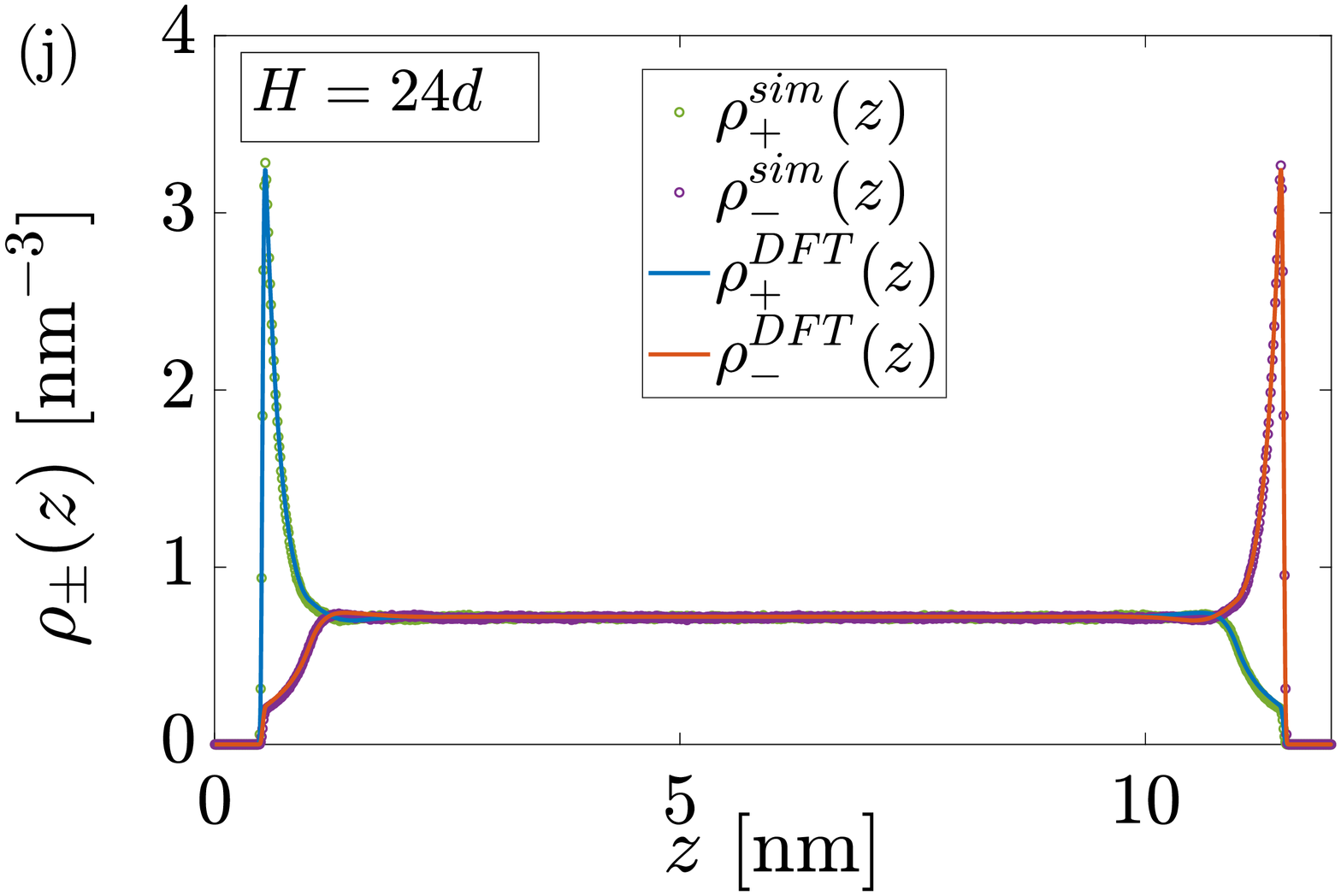}
    \caption{\label{fig:sets}Density profiles of the sets specified in the main text. The solid lines represent the DFT calculations and the symbols the BD simulations. The left electrodes placed at $z=0$ has a negative surface potential attracting cations (blue line for DFT and green circles for BD simulations) and repelling the anions (solid orange line for DFT and purple circles for BD simulations)  and the right electrode places at $z=H$ has a positive potential. In (a) and (b) we change the concentrations w.r.t. reference case, in (c) and (d) the size of the cations, in (e) and (f) the potential, in (g) and (h) the valency of the cation, and in (i) and (j) the electrode separation. The $x$-labels are the same for (a)-(h) and are given in (g) and (h).}
\end{figure}


\subsection{Varying ion concentrations}

The first parameter to be varied is
    the concentration of the electrolyte in the reservoir,
  from $\rho_r=0.1$\,mol$\cdot$L$^{-1}$ to $\rho_r=5$\,mol$\cdot$L$^{-1}$,
    corresponding to 9 and 252 ion pairs in the BD simulation,
      respectively.
      The latter concentration  is similar to that of the ion concentration in ILs, while the dielectric constant
        and the shapes of the ions are markedly different in ILs.
The results, see Fig.~\ref{fig:sets}(a) and  Fig.~\ref{fig:sets}(b),
  reveal a slight difference at the peaks for both cases.
The lower density shows only one peak,
  so we can consider this to be in the diluted limit.
The higher density shows strong oscillations,
  due to the layering commonly found at high packing fractions,
    like the current $\eta=0.394$.
The wavelength and amplitude of the oscillations are well captured with DFT,
  as indicated by the good agreement between the two methods.
Nevertheless,
  there are small discrepancies in the heights of the peaks and valleys.
These are mainly due to
  the treatment of the hard-core potential
    by the FMT part of the functional \cite{Davidchack16}
  and the difference in the representation of the repulsive interactions
    as WCA in BD and as hard spheres in DFT. 


\subsection{Differently-sized ions \label{set2}}

Figure~\ref{fig:sets}(c) and (d) show results for differently-sized ions,
  with cations twice the diameter of anions, $d_+ = 2 d_- = 0.5$\,nm.
The difference between the two plots is
  in the ensemble being used to control the ion number densities in the slit.
Because the anions are smaller than the cations,
  they both come closer to the electrode
    --~thereby lowering their electrostatic energy~--
  and pack at a higher density.
Consequently,
    as illustrated in Fig.~\ref{fig:sets}(c),
  when using DFT 
      to impose reservoir concentration of $\rho_r=1$\,mol$\cdot$L$^{-1}$ for both ions,
    the heights of the density peaks adjacent to both electrodes
        become unequal
    at the reference potentials of $\Phi_L=-0.1$\,V and $ \Phi_R=0.1$\,V,
        hence $\Psi=0.2$ V.
From these density profiles,
  the numbers of ions were calculated as $N_+ = 51$ and  $N_- = 55$,
    to the nearest integer,
  and these numbers were used in the BD simulations shown in the same plot.
Figure~\ref{fig:sets}(d) presents results
  for equal numbers of cations and anions in the slit,
    taken as the average of the two previous values: $N_\pm = 53$.
To obtain the desired numbers of ions of each type in DFT,
  the potential of one electrode was fitted,
      at constant potential difference $\Psi$ 
    to arrive at $\Phi_L=-0.1116$ V and $\Phi_R=0.0884$ V.
The density peaks at both electrodes now resemble each other.
Note the asymmetry in both plots in
  the density of the cations at the positive electrode
    versus that of the anions at the negative electrode,
  both in their distance from the wall
      relative to the other ion at the same wall
    and the distance to the electrode before reaching
      the constant density plateau in the center of the slit.
For both equal and unequal numbers of ions,
  BD and DFT show good agreement in the density profiles.


\subsection{Changing the Surface Potential}\label{sec:changing_potential}

Figure~\ref{fig:sets} (e) and (f) present the comparison
  for smaller $\Psi=0.02$\,V and larger $\Psi=2$\,V
    potential differences with respect to the reference system.
Because convergence of the density profile at the lower potential
    required a very long BD production run,
  the simulation was performed instead
    using Newtonian mechanics in combination with a Nos\'{e}-Hoover thermostat
      \cite{AllenTildesley,FrenkelSmit}.
The thermostat works by rescaling the velocities of all ions at every time step,
  in such a ways as to recover the correct mean kinetic energy
    and kinetic energy fluctuations at the desired temperature,
  and therefore samples the Boltzmann equilibrium distribution
    also obtained by the BD simulation.
The thermostatted method samples configuration space more efficiently
    by ignoring the slow Brownian motion,
  which affects the dynamical properties of the system
    but not the thermodynamic properties studied in this work.

Applying a small potential on the electrodes causes a smaller charge density, i.e. the density profiles of the cations and anions are more similar. At high potentials, a large portion of the ions are adsorbed onto the electrodes, causing strong layering.
The agreement between the simulations and DFT
    is excellent for the smaller potentials,
  while small deviations are observed
    for the larger potential near the second peak.
The latter are a result of strong packing,
  where, as mentioned before,
    the WCA potential differs from
      the FMT approximation to the hard sphere potential in DFT.

\subsection{Different Valencies}

In Fig.~\ref{fig:sets}(g) and (h),
  the valency of the cation is doubled to twice that of the anion,
    $Z_+ = -2 Z_- = 2$.
Charge neutrality of the reservoir, $\sum_j Z_j \rho_{r,j} = 0$,
  implies that the anion concentration in the reservoir
    must be double the cation concentration. Here we choose the cation reservoir concentration
       to be the same as in the reference system, i.e. $2 \rho_{r,+} = \rho_{r,-} = 2$\,mol$\cdot$L$^{-1}$.
Because the symmetry between cations and anions is broken,
  ions exchange with the reservoir resulted
    in distinct number of ions
      in the DFT calculations of Fig.~\ref{fig:sets}(g).
The corresponding BD simulations were based on $N_+ = 35$ and $N_- = 67$,
  following the procedure outlined in Sec.~\ref{set2}.
The simulations in Fig.~\ref{fig:sets}(h)
  impose charge neutrality in the slit, $2 N_+ = N_- = 70$.
The situation is realized in DFT at the electrode potentials
  $\Phi_L = -0.091$\,V and $\Phi_R = 0.109$V at the left and right electrode, respectively. 
The most interesting difference between the reference system and this case is the little hump in the anion density profiles around $z=1$ nm. This effect is referred to as overscreening \cite{Bazant_2011}, where one finds a negatively charged layer of anions next to the positively charged first layer of cations adjacent to the cathode.
Again, excellent agreement is observed between DFT and BD.


\subsection{Slit width}

Lastly,
  the distance between the electrodes $H$ was varied 
  from $H=1.5$\,nm in Fig.~\ref{fig:sets}(i)
    to $H=12$\,nm in Fig.~\ref{fig:sets}(h).
Because a very long production run was required in the BD simulations,
  these were performed using the Nos\'{e}-Hoover thermostat
      rather than by Langevin Dynamics,
    as explained in subsection~\ref{sec:changing_potential}. 
At the lower slit width the EDLs overlap substantially,
  while for the large width the electrolyte acquires
    the flat distribution of a bulk fluid in the middle of the system.
The DFT and BD results are again in excellent agreement.


\subsection{Capacitance\label{sec:Cap}}

\begin{figure}
    \centering
           \includegraphics[width = 0.49\textwidth]
           {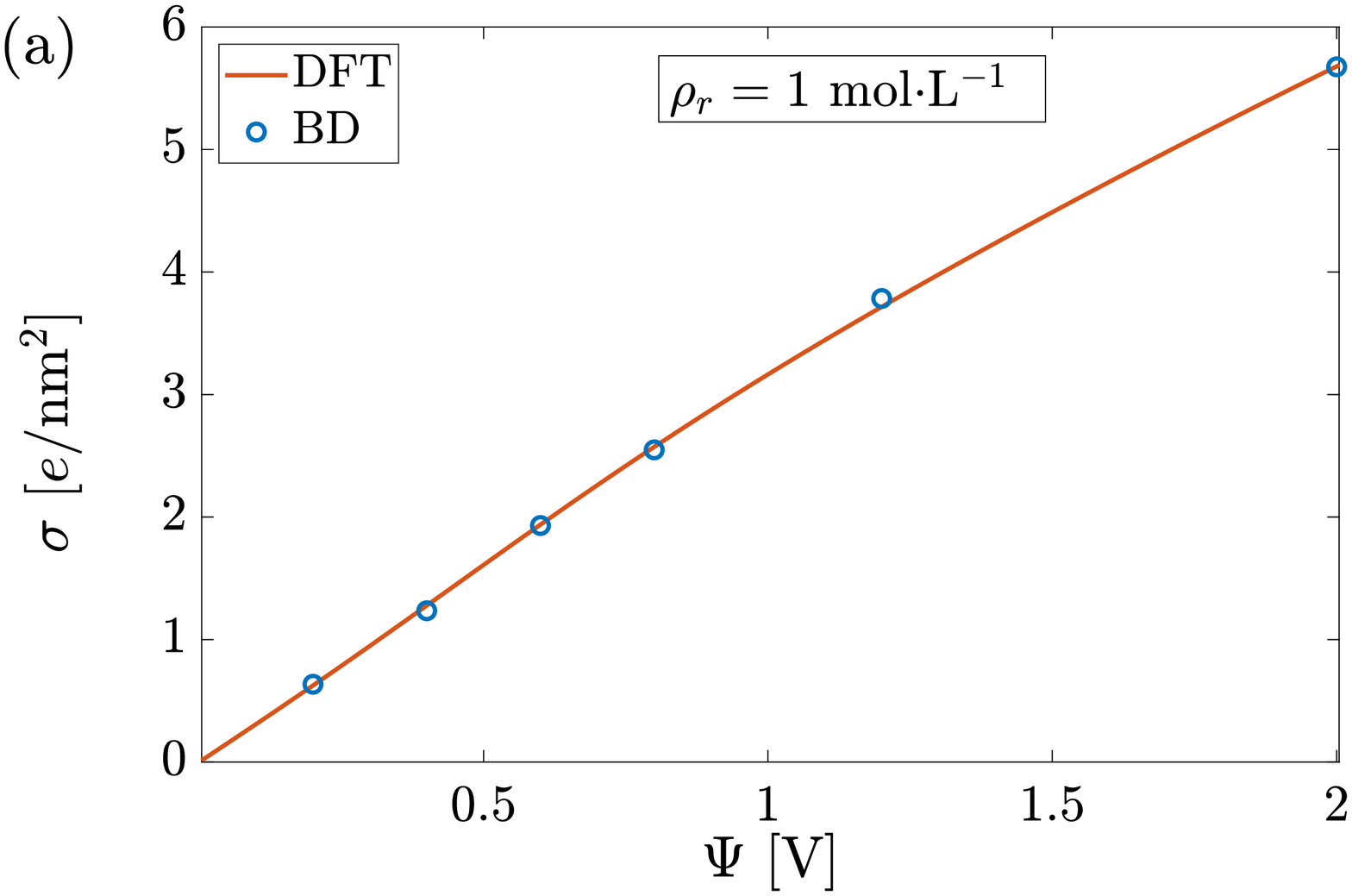}  
           \includegraphics[width = 0.49\textwidth]
           {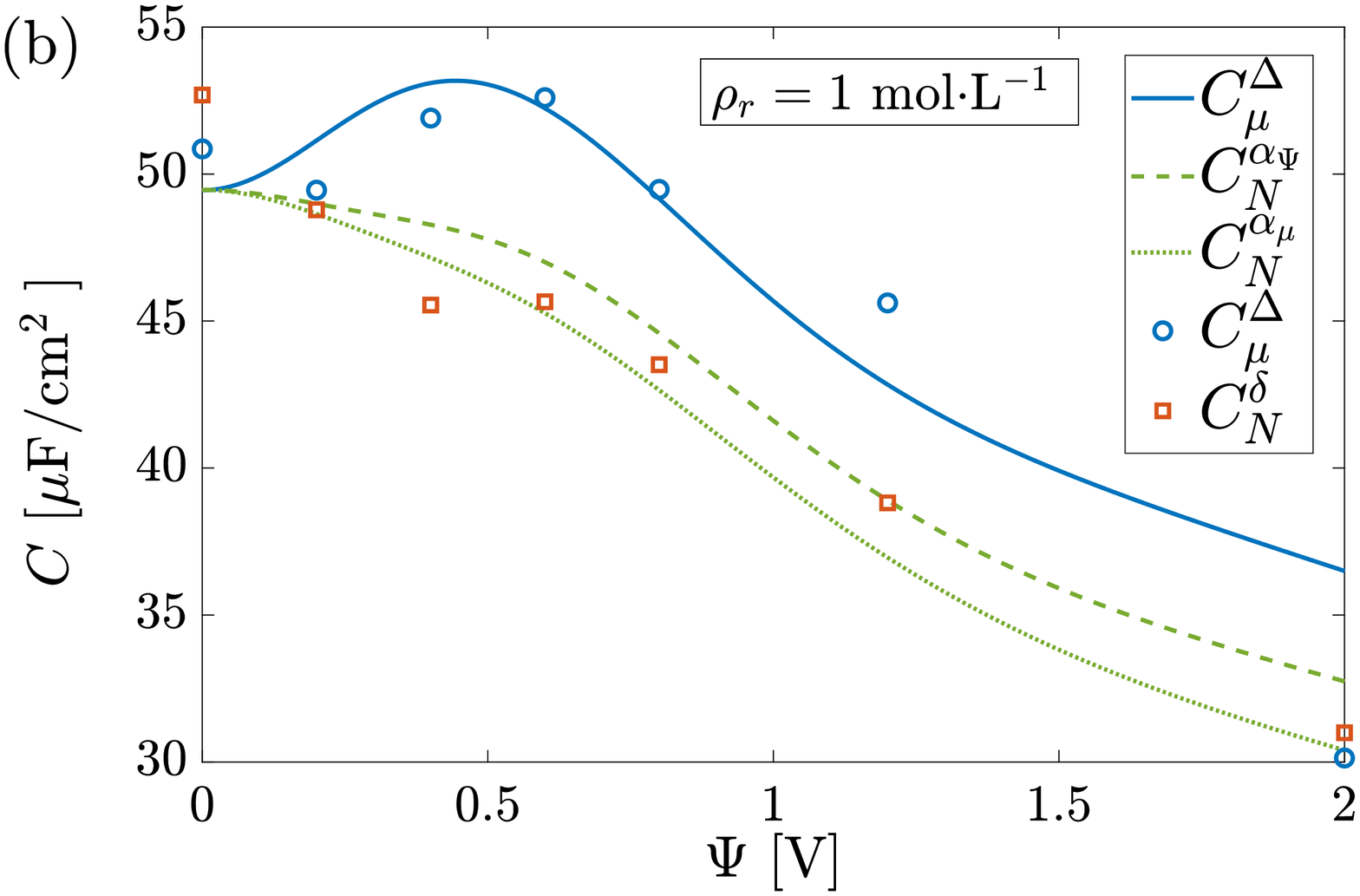} 
    \caption{(a) The average surface charge on the electrodes
        $( \sigma_L - \sigma_R ) / 2$
      and (b) the corresponding differential capacitance,
        as function of the potential difference between the electrodes,
        for a system in thermal equilibrium with
          a reservoir at a salt concentration of 1\,mol$\cdot$L$^{-1}$,
        by DFT and BD calculations.
      The number of ion pairs in the BD simulations
          was determined by DFT,
        and varies with the potential.
    \label{fig:C_1M}
  }
\end{figure}

The capacitance was calculated
    as a function of the surface potential difference $\Psi$,
  at both a constant reservoir concentration and
  at constant number of ion pairs in the slit.
In the former case,
    which comes naturally to DFT,
  the DFT calculations at a concentration in the reservoir of $\rho_r=1$\,mol$\cdot$L$^{-1}$
    were used to determine the numbers of ions in the BD simulations.
In the latter case,
    which comes naturally to BD,
  the number of ion pairs was fixed at $N = 156$
    and the concentration in the reservoir was varied
      to reach the desired number of ions in DFT.
The charge on the electrode surface in the former case
    is shown in Fig.~\ref{fig:C_1M}(a),
  where the line represents the DFT calculations
    and the markers the BD simulations.
Note that each simulation,
    although performed at constant numbers of ions,
  belongs to the same chemical potential. 
As expected,
  the charge on the wall and the number of ions in the slit
    increase with the potential difference between the electrodes.
The two methods are in good agreement.
The corresponding capacitance is presented in Fig.~\ref{fig:C_1M}(b),
  where several calculation methods have been used.
The blue solid line and the blue circles represent $C_\mu^\Delta$ Eq.~\eqref{Eq:CDmu} using the data in Fig.~\ref{fig:C_1M}(a). 
Because the number of time-consuming BD simulations
    is necessarily low,
  the numerical derivative is limited in its accuracy, especially for the last data point at $\Psi=2$ V.
Nevertheless,  the agreement is satisfactory and
   both methods yield similar camel-shaped curves \cite{Korny}. 
Also shown in Fig.~\ref{fig:C_1M}(b) are calculations of $C_N$,
  where it should be emphasized that
    $N$ is not constant across the plot but varies with $\Psi$.
The capacitance $C_N^\delta$ (orange squares)
    is based on the charge fluctuations in the BD simulations, given in Eq.~\eqref{Eq:CN_fluc} where also $C_0$ appears. The value for $C_0$, which depends neither on the number of particles nor on the potential difference, is found to be
      $C_0=17.6$\,$\mu$F\,cm$^{-2}$ (see Appendix~\ref{App:calc_cap}). The DFT calculations of $C_N^{\alpha_\mu}$ (green dotted line) and $C_N^{\alpha_\Psi}$ (green dashed line)
    are based on the relations in Eq.~\eqref{Eq:CmuCN}.
The approximation made in Sec.~\ref{sec:DFT},
  namely the assumed dependence of
    the direct correlation function $c^{MSA}$ in Eq.~\eqref{Eq:cMSA}
      on the chemical potential,
  resurfaces at this point.
The adsorption $\Gamma$ can be obtained
  either from Eq.~\eqref{Eq:Gamma}
  or from the derivative of Eq.~\eqref{Eq:Lipmann}
    with respect to the chemical potential,
      as derived in Appendix~\ref{App:Gam_incon}.
In the latter case,
  the derivative of $\mathcal{F}^{ES}_{ex}[\{\rho\}]$ does not vanish,
  though in principle it should have.
The capacitance $C_N$ can therefore be calculated from $C_\mu$
  using either $\alpha_\Psi$ or $\alpha_\mu$,
  where the expression for $\Gamma$ in Eq.~\eqref{Eq:Gamma}
    was used to calculate $\alpha_\Psi$.
Note, however, that the calculation of the surface charge density $\sigma$
    \textit{is} consistent by construction,
      since charge neutrality is imposed.

As expected from Eq.~\eqref{Eq:CapmuN},
  in both cases $C_N$ is smaller than $C_\mu$.
Both DFT calculations are in reasonable agreement with the BD results;
  notably, all three show a bell-shaped curve.
The plot shows a substantial difference between $C_N$ and $C_\mu$,
  and although the various calculations
    do not exactly match quantitatively,
  they agree reasonably well and support the qualitative difference. The reason for the rather large difference between $C_\mu$ and $C_N$ is due to the small electrode-electrode separation, where the region of the EDLs contribute substantially to the total number of ions in the system. In the limit where the separation between the electrodes is infinite, the difference between $C_\mu$ and $C_N$ disappears.

\begin{figure}
    \centering
    \includegraphics[width = 0.6\textwidth]{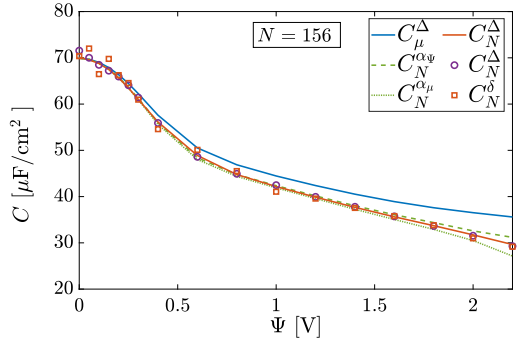}
    \caption{The differential capacitance
      as a function of potential difference between the electrodes,
          using BD and DFT,
        for a system containing 156 ion pairs.
    Because the number of ion pairs is fixed,
      their chemical potential varies
        with the electrostatic potential difference.
    \label{fig:C_156N}
    }
\end{figure}

Lastly, we consider the system with a fixed number of ion pairs,
    $N = 156$,
  and vary the surface potential difference $\Psi$. Shown in Fig.~\ref{fig:C_156N} are the capacitances using the same colour and line coding as in Fig.~\ref{fig:C_1M}, with the addition of an orange curve for $C_N^\Delta$ using DFT. For $\Psi = 0$ V to $\Psi = 0.3$ V, the simulations were run for 800\,ns, treating the first 200\,ns as equilibration phase, since for lower potential differences the simulations required a long production run for the capacitances to converge.
Both differential capacitances are bell-shaped.
The agreement between simulations and DFT is remarkably good, and compared to the results in Fig.~\ref{fig:C_1M}(b), there is little to no qualitative difference between $C_N$ and $C_\mu$. The small qualitative difference, especially at small potential differences, is mainly due to the relative large number of ions in the system and therefore a corresponding large reservoir concentration. For comparison, the number of ions at $\Psi=0.2$ V in Fig.~\ref{fig:C_1M} is 51, while only at $\Psi=2$ V it is 156. The camel shaped curve in $C_\mu$ is only existent for small reservoir concentrations $\rho_r<1.5$\,mol$\cdot$L$^{-1}$ and bell shaped otherwise. Hence, no camel-shaped capacitance curve is observed within these parameters.


\section{Discussions, Conclusions and Outlook}

We presented ionic density profiles for a broad range of parameters applicable to aqueous electrolytes confined between a planar cathode and anode, and found very good agreement between results from DFT calculations and BD simulations. Both methods were also used to calculate differential capacitances, either $C_\mu$ at constant ionic chemical potential $\mu$ or $C_N$ at constant number $N$ of ions, via several routes. For a fixed chemical potential of mono-valent ions, at which the ionic reservoir concentration equals $\rho_r=1$\,mol$\cdot$L$^{-1}$, the capacitance curves obtained from DFT and BD are overall in good agreement. The DFT prediction for the capacitance at fixed $N$, however, gave two somewhat different results due to the approximation for the electrostatic part of the employed functional. Nevertheless, the DFT predictions bracket those of the BD simulations, except at potential differences between cathode and anode below 0.3 V where the simulations were extremely slow. Interestingly, the qualitative difference between $C_\mu$ and $C_N$ is substantial, where $C_\mu$ is camel shaped and $C_N$ is bell shaped. This has to do with the nonlinear relation between $\mu$ and $N.$ 
Furthermore, $C_\mu$ is found from a linear cut through the landscape in the three-dimensional space spanned by $\{\sigma,\Psi,\mu\}$, whereas $C_N$ is the result of a non-trivial path through this landscape. We also considered capacitance curves at constant numbers of ions, $N=156$, and found excellent agreement between DFT and BD simulations. In this case there is no qualitative difference between $C_\mu$ and $C_N$. Let us stress the time it takes to obtain the results from DFT and BD simulations. A typical BD simulation of a state point took 2 to 3 days on 32 cores, whereas the DFT calculations took not even 2 seconds on a regular laptop, which amounts to a difference of about 7 orders of magnitude. The accuracy that is lost by applying DFT on these systems is very small, as we have shown throughout this manuscript. 

We conclude that with DFT one can obtain the same accuracy in \textit{structural and thermodynamic} quantities as in BD simulations, at least for aqueous systems.  This allows one to explore parameter space much more effectively and to study the properties of these systems thoroughly. A drawback of DFT is that it gives only an equilibrium description of the system, whereas BD simulations also provide the dynamics. We furthermore conclude that one needs to be careful and specify the differential capacitance that is being studied, e.g. $C_\mu$ or $C_N$, because they can differ both qualitatively and quantitatively. Also the natural choice changes depending on the method employed i.e. $C_\mu$ for DFT and $C_N$ for BD, meaning the direct comparison of results from different methods is not straightforward.

A natural next step will be to divert from aqueous systems, to study systems with a lower dielectric constant. An interesting direction will be to study room temperature ionic liquid (ILs).
Although the concentration of $\rho_r=5$\,mol$\cdot$L$^{-1}$ in Fig.~\ref{fig:sets}(b) is comparable to that of ionic liquids, the dielectric constant here is considerably higher
  due to the solvent.
It is not sufficient to simply reduce the dielectric constant and redo the calculations, since the electrostatic correlations become much stronger at the low dielectric constants of ILs: a cation-anion pair of sub-nm diameter will bind at contact by Coulombic attractions of several tens of $k_B T$.  It is therefore not evident whether DFT or BD simulation will work in this regime.  Moreover, the ionic shape in ILs is often non-spherical and needs to be accounted for in DFT. Interestingly, there have been developments in DFT to account for chain-like ions and molecules \cite{Yu_2002}, and these have been applied  to some extent to study ILs \cite{Forsman_2011,Fedorov_2014,Henderson_2011,Jiang_2011,Yang_2020,Shen_2020}. Besides chain-like ions, another approach to implement shape and polarizability is via molecular DFT \cite{Levesque_2012,Jeanmairet_2013,Ding_2017,Jeanmairet_2019}. Although molecular DFT has been mostly applied to model water, it might prove worthwhile to use this approach for ionic liquids in the future.
Not only DFT is challenging at lower dielectric constant, but also simulations become much more challenging due to clustering that occurs at low dielectric constants as a result of the stronger electrostatic interactions. This leads to longer simulation times, which were already non-negligible in the aqueous systems. Hence, a proper functional can provide the means to study ILs and electrolytes at low dielectric media effectively.

A different line of investigation would be the study of the differences of the differential capacitance $C_N$ and $C_\mu$. Until now the distinction has not often been made explicitly and further studies are needed to map out the properties and relations between both.

\section*{Acknowledgement}
We would like to thank our project partners at JNCASR in Bangalore,  in particular, Prof. S. Balasubramanian and Nikhil V. S. Avula for fruitful discussions.

This work forms part of the D-ITP consortium and the Data-driven science for smart and sustainable energy research program, with project number 16DDS014. Both programs are from the Netherlands Organisation for Scientific Research (NWO) that is funded by the Dutch Ministry of Education, Culture and Science (OCW).

\bibliographystyle{unsrt} 
\bibliography{references}   


\appendix

\section{Thermodynamics Derivation \label{App:Thermo}}
For details see Ref.\cite{Roijstat}.
The differential for surface grand potential $\gamma$ (or also called the surface tension) can be derived from taking the differential of $\Omega=-pV+\gamma A$ and equating it with Eq.~\eqref{Eq:dOm}, i.e.
\begin{align}
    \mathrm{d}\Omega&=-p\mathrm{d}V-V\mathrm{d}p+\gamma \mathrm{d}A+A\mathrm{d}\gamma\\
    &=-S\mathrm{d}T-p\mathrm{d}V-\sum_j N_j \mathrm{d}\mu_j-A\sigma_L\mathrm{d}\Phi_L-A\sigma_R\mathrm{d}\Phi_R+\gamma \mathrm{d}A\green{-}f\mathrm{d}H.
\end{align}
One now needs to separate the volumetric bulk terms from the surface terms. For convenience, let us define $N_j=V\rho_{r,j}+A\Gamma_j$ and $S=Vs_r+As_s$, where $\rho_{r,j}$ and $s_r$ denote the particle density and the entropy density in the reservoir, respectively, while $\Gamma_j$ denotes the adsorption of species $j$ and $s_s$ the areal excess entropy.
This separation into volumetric and surface terms allows us to properly gather the volume terms on the left and the surface terms on the right of the equation, i.e.
\begin{align}
    -V\Big(\mathrm{d}p-s_r\mathrm{d}T-\sum_j \rho_{r,j}\mathrm{d}\mu_j\Big)=A\Big(-\mathrm{d}\gamma-s_s\mathrm{d}T-\sum_j \Gamma_j\mathrm{d}\mu_j-\sigma_L\mathrm{d}\Phi_L-\sigma_R\mathrm{d}\Phi_R\green{-}f\mathrm{d}H\Big).
\end{align}
Given that the surface has no influence on the volume term, both sides vanish and we obtain both the Gibbs-Duhem as well as the full Lipmann equation
\begin{align}
    \mathrm{d}p&=s_r\mathrm{d}T+\sum_j\rho_{r,j}\mathrm{d}\mu_j,\\
    \mathrm{d}\gamma&=-s_s\mathrm{d}T-\sum_j\Gamma_j\mathrm{d}\mu_j-\sigma_L\mathrm{d}\Phi_L-\sigma_R\mathrm{d}\Phi_R\green{-}f\mathrm{d}H.
\end{align}
The systems that we considered when calculating the capacitances were at constant temperature $T$ and constant electrode separation $H$. Moreover, in those calculations we considered the RPM such that $\Gamma_+=\Gamma_-\equiv\Gamma/2$ and $\mathrm{d}\mu_+=\mathrm{d}\mu_-\equiv\mathrm{d}\mu$ and we symmetrized the surface potentials so that $\mathrm{d}\Phi_L=-\mathrm{d}\Phi_R\equiv\mathrm{d}\Psi/2$ and $\sigma_L=\sigma_R\equiv\sigma$. The Lipmann equation in this situation simplifies to
\begin{align}\label{Eq:app_gamma}
    \mathrm{d}\gamma=-\Gamma\mathrm{d}\mu-\sigma\mathrm{d}\Psi.
\end{align}
Within the $\Omega$ ensemble $\sigma(\mu,\Psi)$ is a function of $\mu$ and $\Psi$, hence
\begin{align*}
    \mathrm{d}\sigma=\left(\frac{\partial \sigma}{\partial \Psi}\right)_\mu\mathrm{d}\Psi+\left(\frac{\partial \sigma}{\partial \mu}\right)_\Psi\mathrm{d}\mu.
\end{align*}
Because $N(\mu)$ is a function of $\mu$, we therefore find that
\begin{align}\label{Eq:CmuCPsipre}
    \left(\frac{\partial \sigma}{\partial \Psi}\right)_N=\left(\frac{\partial \sigma}{\partial \Psi}\right)_\mu+\left(\frac{\partial \sigma}{\partial \mu}\right)_\Psi\left(\frac{\partial \mu}{\partial \Psi}\right)_N,
\end{align}
which can be rewritten using the Maxwell relation $\left(\frac{\partial \sigma}{\partial \mu}\right)_\Psi=\left(\frac{\partial \Gamma}{\partial \Psi}\right)_\mu$, that can be obtained from Eq.~\eqref{Eq:app_gamma}, and the identity $\left(\frac{\partial \mu}{\partial \Psi}\right)_N=-\left(\frac{\partial \mu}{\partial N}\right)_\Psi\left(\frac{\partial N}{\partial \Psi}\right)_\mu$. Using the differential capacitances from Eq.~\eqref{Eq:CapmuN} and introducing the compressibilities
\begin{align}
    \chi_\sigma=\left(\frac{\partial N}{\partial \mu}\right)_\sigma, & &     \chi_\Psi=\left(\frac{\partial \Gamma}{\partial \mu}\right)_\Psi,
    \end{align}
    and the Maxwell relation
    \begin{align}
    \alpha_\mu&:=\left(\frac{\partial N}{\partial \Psi}\right)_\mu=\left(\frac{\partial \sigma}{\partial \mu}\right)_\psi=:\alpha_\Psi,
\end{align}
one can rewrite Eq.~\eqref{Eq:CmuCPsipre} as
\begin{align}
   C_\mu-C_N=\frac{\alpha_\Psi^2}{\chi_\Psi} \geq 0.
\end{align}
Using the same relations, one can show that
\begin{align}
    \frac{C_\mu}{C_N}&=\frac{\left(\frac{\partial \sigma}{\partial \Psi}\right)_\mu}{\left(\frac{\partial \sigma}{\partial \Psi}\right)_N}=\frac{-\left(\frac{\partial \sigma}{\partial \mu}\right)_\Psi\left(\frac{\partial \mu}{\partial \Psi}\right)_\sigma}{-\left(\frac{\partial \sigma}{\partial N}\right)_\Psi\left(\frac{\partial N}{\partial \Psi}\right)_\sigma}=\frac{\left(\frac{\partial \sigma}{\partial \mu}\right)_\Psi\left(\frac{\partial N}{\partial \sigma}\right)_\Psi}{\left(\frac{\partial \Psi}{\partial \mu}\right)_\sigma\left(\frac{\partial N}{\partial \Psi}\right)_\sigma}=\frac{\left(\frac{\partial N}{\partial \mu}\right)_\Psi}{\left(\frac{\partial N}{\partial \mu}\right)_\sigma}\\
    &=\frac{\chi_\Psi}{\chi_\sigma}\geq 1.
\end{align}
These relations allows us to relate the differential capacitance obtained from simulations $C_N$ to the differential capacitance from DFT $C_\mu$. Within DFT one can calculate $C_\mu$, $\chi_\Psi$ and $\alpha_\Psi$, which through Eq.~\eqref{Eq:CmuCN} gives access to $C_N$. Notice the similarity with the relations between the heat capacity at constant pressure $c_p$ and the heat capacity at constant volume $c_V$ (see e.g. the thermodynamics book \cite{Blundell}). Although this derivation was done for the symmetric RPM, one can generalize these equations for any system. This is necessary when considering unequal ion sizes/valencies, but also when the potential on both electrodes differ. Hence, in general one needs to consider both electrodes separately.


\section{Calculation of the Capacitance\label{App:calc_cap}}

\begin{figure} 
    \centering
 \includegraphics[width=0.45\textwidth]
      {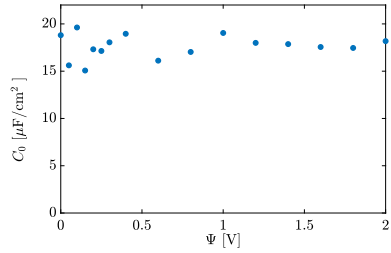} \\
    \caption{The difference in BD between the capacitance $C_N$
      as obtained by numerical differentiation
          of the surface charge with potential, see Eq.~\eqref{Eq:CapmuN},
        and as obtained from the thermal charge fluctuations only,
          see Eq.~\eqref{Eq:CN_fluc},
        confirms that the constant $C_0$ in the latter expression
          is indeed independent of the potential difference.}
    \label{fig:C_0}
\end{figure}

The calculation of a differential capacitance $C_N$ in BD
    using the fluctuation expression of Eq.~\eqref{Eq:CN_fluc}
  requires the evaluation of the constant $C_0$
    accounting for the neglected thermal charge fluctuations
      around the idealized charges calculated by CPM.
Because $C_0$ is a property of the electrodes
    that depends neither on the number of ions nor on surface potential,
  its value was determined as
    the difference between
      the $C_N$ obtained from the conventional charge-potential relation,
          see Eq.~\ref{Eq:CapmuN},
      and that obtained from the fluctuating contribution
          in Eq.~\eqref{Eq:CN_fluc} for $C_0 = 0$.
This difference,
    plotted in Fig.~\ref{fig:C_0} for a range of potentials,
  appears indeed to be a constant, with a value of $C_0=17.6$\,$\mu$F\,cm$^{-2}$.
The noise in the data is higher at low potentials,
  because the simulations converge more slowly at low potentials
    as well as due to the increased signal-to-noise ratio
      at the smaller step sizes in numerical differentiation.


\section{Adsorption Inconsistency}\label{App:Gam_incon}

\begin{figure}
    \centering
     \begin{tabular}{c c}
           \includegraphics[width=0.46\textwidth]{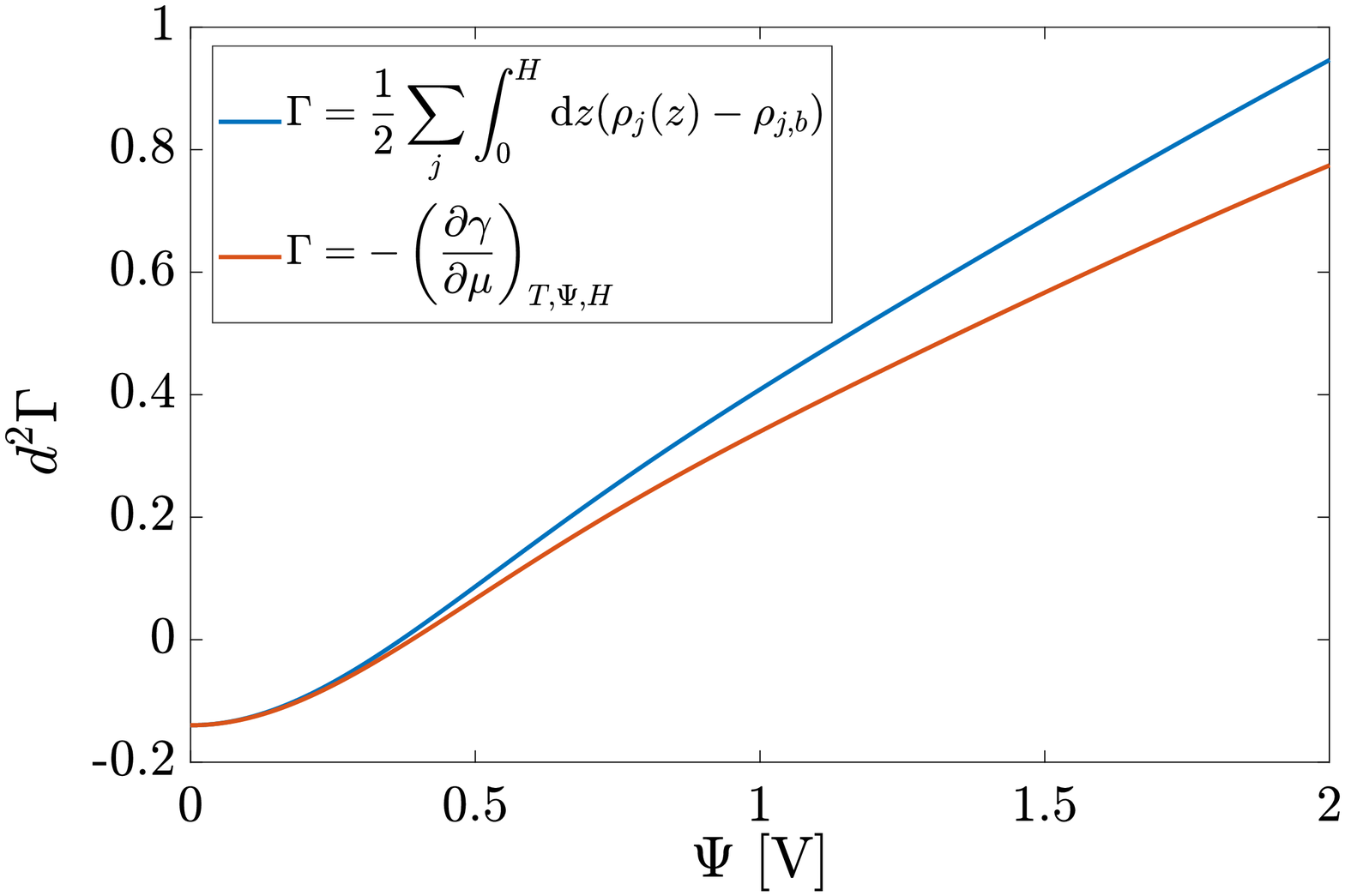}  &    \includegraphics[width=0.46\textwidth]{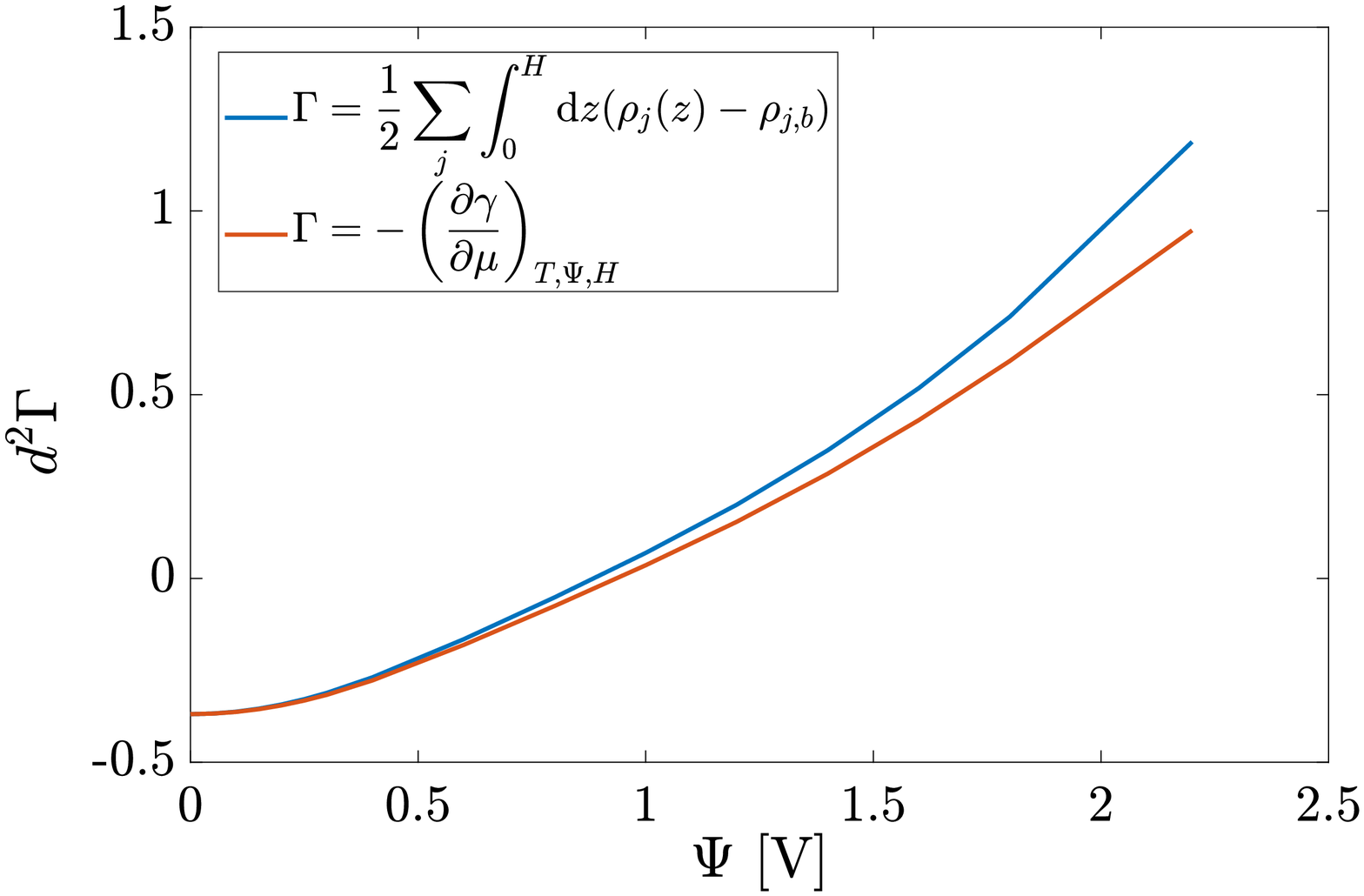} \\
           (a) & (b) \\
    \end{tabular}
    \caption{The inconsistency in the adsorption at (a) a fixed reservoir concentration of $\rho_r=1$\,mol$\cdot$L$^{-1}$ and at (b) a fixed number of ion pairs $N=156$ (b).}
    \label{fig:Gam_incons}
\end{figure}

The adsorption can be calculated via the two routes
\begin{align}
\Gamma=-\left(\frac{\partial \gamma}{\partial \mu}\right)_{T,\Psi,H}, & & \mathrm{and} & & \Gamma=\frac{1}{2}\sum_{j=\pm}\int_0^H\left(\rho_j(z)-\rho_{j,b})\right),
\end{align}
which are both plotted in Fig.~\ref{fig:Gam_incons} at a reservoir concentration of $\rho_r=1$\,mol$\cdot$L$^{-1}$. As a short note, any functional that is based upon a bulk expansion like the one we use for the electrostatics suffers from this inconsistency.

\section{Constant potential versus fixed charge\label{App:CMP_vs_FCP}}

\begin{figure}
    \centering
    \includegraphics[width = 0.6\textwidth]{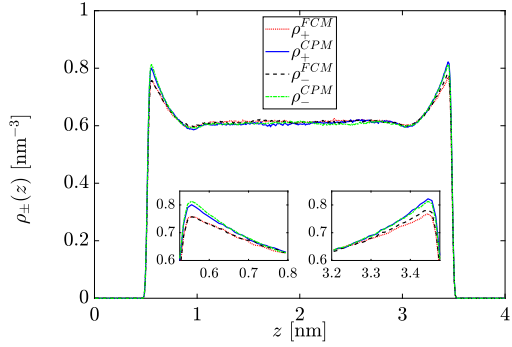}
    \caption{
      Number densities of cations, $\rho_+$, and anions, $\rho_-$,
        for the fixed charged method (FCM)
            at vanishing surface charge on the electrodes and
          the constant potential method (CPM)
            at vanishing potential difference between the electrodes,
        for capacitors in equilibrium with a 1\,mol$\cdot$L$^{-1}$ salt concentration.
      The insets show enlargements of
        the density peaks adjacent to the electrode surface.
    \label{fig:FCM_vs_CPM}
    }
\end{figure}

The constant potential method (CPM) and fixed charge method (FCM)
  were compared by running simulations at zero voltage and zero charge,
    respectively.
Based on DFT calculations,
  the equilibrium with a 1\,mol$\cdot$L$^{-1}$ reservoir
    results in 47 ion pairs in the simulated slit.
 In the BD simulations
   the total charges on the left and right electrodes fluctuate
       around averages of $\pm 7\,n$C$\cdot$cm$^{-2}$,
     which is less then 1\% of their standard deviations
       of 0.8\,$\mu$C$\cdot$cm$^{-2}$.
Hence,  the mean total charge on the electrodes is essentially zero. It should be noted that  the \texttt{slab} option in LAMMPS can handle non-neutral systems \cite{Yeh_and_Berkowitz, Ballenegger}.
The ionic number densities in both
  CPM and FCM simulations are similar,
    see Fig.~\ref{fig:FCM_vs_CPM}.
The slightly higher density peaks near the electrodes for CPM
  are probably caused by the ions inducing a mirror charge in the electrode
    and being attracted by that mirror charge;
  this effect evidently does not occur at fixed wall charges.



\end{document}